\documentclass[journal, 12pt,draftcls,onecolumn,twoside]{IEEEtran}
\ifCLASSINFOpdf
\else
\fi
\usepackage{url}
\usepackage[utf8]{inputenc} 
\usepackage[T1]{fontenc}
\usepackage{url}
\usepackage{ifthen}
\usepackage{cite}
\usepackage[cmex10]{amsmath} % Use the [cmex10] option to ensure complicance
\usepackage{cite}
\ifCLASSINFOpdf
 
\else
   \usepackage[dvips]{graphicx}
 \fi
\usepackage[cmex10]{amsmath}
\usepackage{amssymb}
\usepackage{stfloats}
\usepackage{amsthm} 
\usepackage{wrapfig}
\usepackage{amssymb}
\usepackage{latexsym}
\usepackage{bm, bbm} 
\usepackage{epsfig}
\usepackage{graphicx}
\usepackage{epsfig}
\usepackage{multicol}
\usepackage{psfrag}
\usepackage{float}
\usepackage{cite}
\usepackage{subfigure}
\usepackage{color}
\usepackage{balance}

\newcommand{\Ftwo}{\mathbb{F}_2} 
\newcommand{\code}[1]{\mathcal{#1}}
\newcommand{\matr}[1]{{#1}}

\newcommand{\vc}{\vect{c}}
\newcommand{\tr}{\mathsf{T}}
\newcommand{\vect}[1]{\mathbf{#1}}
\newcommand{\set}[1]{\mathcal{#1}}
\newcommand{\dmins}{d_{\mathrm{min}}}
\newcommand{\defeq}{\triangleq}
\newcommand{\cC}{\mathcal{C}}
\def\girth{{\rm girth}}
\def\diag{{\rm diag}}
\def\rank{{\rm rank}}

\newtheorem{lemma}{Lemma}
\newtheorem{theorem}{Theorem}

\newtheorem{corollary}{Corollary}

\theoremstyle{plain}

\newtheorem{PreDefinition}[lemma]{{\textbf{Definition}}}
    {\begin{PreDefinition}}{\hfill$\square$\end{PreDefinition}}

\theoremstyle{plain}

\newtheorem{PreRemark}[lemma]{{\textbf{Remark}}}
  \newenvironment{remark}%
    {\begin{PreRemark}\upshape}{\hfill$\square$\end{PreRemark}}

\newtheorem{PreExample}{{\textbf{Example}}}
  \newenvironment{example}%
    {\begin{PreExample}\upshape}{\hfill$\square$\end{PreExample}}

\newcommand{\vertsize}{{n_c}}
\newcommand{\horsize}{{n_v}}

\newcommand{\eg}{\emph{e.g.}}

\newcommand{\ie}{\emph{i.e.}}

\newcommand{\matrH}{\matr{H}}

\newcommand{\shortFtwoxmodN}{\Ftwo^{\langle N \rangle}[x]}

\newcommand{\vs}{\vect{s}}

\long\def\symbolfootnote[#1]#2{\begingroup%
\def\thefootnote{\fnsymbol{footnote}}\footnote[#1]{#2}\endgroup} 

\usepackage{xcolor}

\usepackage{color,soul}

\definecolor{lightblue}{rgb}{.90,.95,1}
\sethlcolor{yellow}

\usepackage{psfrag}

\usepackage[outdir=./]{epstopdf}

\begin{document}
\title{Generalized Quasi-Cyclic LDPC Codes:\\ Design and Efficient Encoding}

%\author{David~G.~M.~Mitchell,~\IEEEmembership{Senior Member,~IEEE,} Pablo M. Olmos,~\IEEEmembership{Member,~IEEE,}\\Michael~Lentmaier,~\IEEEmembership{Senior Member,~IEEE,} and Daniel~J.~Costello,~Jr.,~\IEEEmembership{Life Fellow,~IEEE}% <-this % stops a space
\author{Roxana~Smarandache, David G. M. Mitchell, and Anthony G\'omez-Fonseca% <-this % stops a space

        \thanks{This material is based upon work supported in part by the
National Science Foundation under Grant No. CCF-2145917 and fellowships from  GFSD and Kinesis-Fern\'andez Richards. The material in this paper was presented in part at the 2022 IEEE International Symposium on Information Theory.}% <-this % stops a space
\thanks{R.~Smarandache is with the Departments of Mathematics and Electrical Engineering, University of Notre Dame, Notre Dame, IN 46556, USA (e-mail: rsmarand@nd.edu).}% <-this % stops a space
\thanks{D.~G.~M.~Mitchell is with the Klipsch School of Electrical and Computer Engineering, New Mexico State University, Las Cruces, NM, USA (e-mail: dgmm@nmsu.edu).}% <-this % stops a space
\thanks{A. G\'omez-Fonseca is with the Department of Mathematics, University of Notre Dame, Notre Dame, IN 46556, USA (e-mail: agomezfo@nd.edu).}}% <-this % stops a space% <-this % stops a space
\markboth{IEEE Transactions on Information Theory.}{}%
\maketitle

\begin{abstract}
Generalized low-density parity-check (GLDPC) codes, where single parity-check constraints on the code bits are replaced with generalized constraints (an arbitrary linear code), are a promising class of codes for low-latency communication. The block error rate performance of the GLDPC codes, combined with a complementary outer code, has been shown to outperform a variety of state-of-the-art code and decoder designs with suitable lengths and rates for the 5G ultra-reliable low-latency communication (URLLC) regime. A major drawback of these codes is that it is not known how to construct appropriate polynomial matrices to encode them efficiently. In this paper, we analyze practical constructions of quasi-cyclic GLDPC (QC-GLDPC) codes and show how to construct polynomial generator matrices in various forms using minors of the polynomial matrix. The approach can be applied to fully generalized matrices or partially generalized (with mixed constraint node types) to find better performance/rate trade-offs. The resulting encoding matrices are presented in useful forms that facilitate efficient implementation. The rich substructure displayed also provides us with new methods of determining low weight codewords, providing lower and upper bounds on the minimum distance and often giving those of weight equal to the minimum distance. Based on the minors of the polynomial parity-check matrix, we also give a formula for the rank of any parity-check matrix representing a QC-LDPC or QC-GLDPC code, and hence, the dimension of the code. Finally, we show that by applying double graph-liftings, the code parameters can be improved without affecting the ability to obtain a polynomial generator matrix.
\end{abstract}
%quadrature PSK modulation. In~\cite{8532357},  practical construction of quasi-cyclic GLDPC codes were proposed, with the proportion of generalized constraints determined by an asymptotic analysis. In this paper,  we analyze some of these codes, show that they can have large minimum distance and find generator matrices, thus making them practical. As observed in~\cite{8532357},  the optimal proportion, from a
%threshold perspective, is that a 0.75 percent of the check
%nodes should be replaced by GC nodes in the (2, 6)-regular
%and (2,7)-regular cases, while this fraction increases to the 0.8
%in the (2, 15)-regular case. In this paper we should how to obtain QC-codes that satisfy this criterion.
%The approach is based on previous  results~\cite{9834862} and~\cite{9762914} .... \end{abstract}

\section{Introduction}\label{sec:intro}

%\IEEEPARstart{T}{his} 

%***************************************************************************
%\symbolfootnote[0]{This material is based upon work supported by the National Science Foundation under Grant Nos. OIA-1757207 and HRD-1914635. }
%\symbolfootnote[0]{This material is based upon work supported by the National Science Foundation under Grant Nos. CCF-2145917 and CNS-2148358. A. G. F. thanks the support of the GFSD (formerly NPSC) and Kinesis-Fern\'andez Richards fellowships.}
Generalized low-density parity-check (GLDPC) codes, where single parity-check constraints are replaced by a general linear code, have been shown to have several advantages over conventional LDPC codes, including large minimum distance \cite{lz99,bpz99,llpf10}, good iterative decoding performance \cite{lf10,lom18b}, fast decoding convergence speed \cite{mpf15}, and low error floors \cite{lrc08,molc21}. The performance and convergence speed of the BP decoder can be further improved by applying a reinforcement learning approach to optimize the scheduling of messages in the decoder \cite{hm23}. This improved performance from the generalized constraints comes at the cost of reduced coding rate, which is generally not desirable for many communication systems. However, these codes are well suited for next generation machine-to-machine (M2M) type communications and ultra reliable low latency communications (URLLC), which are expected to have low coding rate, e.g., $R=1/12$, and short block lengths ranging from hundreds of bits to one or two thousand \cite{swj+16,urllc2,lom18b}.

Conventional quasi-cyclic LDPC (QC-LDPC) codes  have a highly structured parity-check matrix, which can be composed as an array of circulant matrices. Consequently, they are attractive for implementation purposes since their structure leads to efficiencies in decoder design \cite{wc07}. Quasi-cyclic GLDPC (QC-GLDPC) codes  were proposed in \cite{lrc08}, where irregular protograph-based designs were shown to possess good performance in both the waterfall and error floor regions. In~\cite{lom18b}, a practical construction of QC-GLDPC codes were proposed for URLLC, where the optimal proportion of generalized Hamming constraints (to minimize the gap to capacity) was determined by an asymptotic analysis. The optimal proportions were found to be 0.75\% of the constraint nodes in the $(2, 6)$-regular and $(2,7)$-regular cases, while this proportion increases to the 0.8\% in the $(2, 15)$-regular case. A major drawback of these codes is that it is not known how to leverage the circulant based generalized matrix to construct appropriate polynomial matrices to encode them efficiently.

  %Moreover, QC-LDPC codes can be shown to perform well compared to random LDPC codes for moderate block lengths \cite{klf01,tss+04}.  However, unlike typical members of an asymptotically good protograph-based LDPC code ensemble, the QC sub-ensemble does not have linear distance growth. Indeed, if the protograph base matrix consists of only ones and zeros, then the minimum Hamming distance is bounded above by $(n_c+1)!$, where $n_c$ is the number of check nodes in the protograph, regardless of the lifting factor $N$ \cite{md01} This result was extended to multi-edge protographs in \cite{Smarandache_2012}. %The girth of the Tanner graph is also bounded above as a result of the circulant structure, see (Fossorier 2004) for necessary and sufficient conditions to determine the girth of a Tanner graph that can be efficiently computed using modular arithmetic. 
%Significant effort has been made in the coding theory community to design QC-LDPC code matrices with minimum distance and girth approaching these bounds, see \cite{klf01,tss+04, kncs07,phns13,kb13,msc14} and references therein.

While conventional techniques to obtain generator matrices from LDPC matrices, such as Gauss-Jordan elimination and variants \cite{ru01}, can be applied to obtain generator matrices for QC-LDPC codes, these will typicaly result in dense and unstructured matrices. Instead, one can take advantage of the rich structure of QC-LDPC codes to obtain generator matrices. These include methods to construct a generator in standard form (e.g., \cite{lcz+06}), which allows high throughput systematic encoding but the resulting generator matrix is typically dense.\footnote{Systematic encoding, where the information is directly embedded in the codeword, is often preferred in practice.} Various other methods have been proposed that allow for sparse and structured generator matrices for QC-LDPC codes, where there are specific constraints placed on the parity-check matrix \cite{baldi2011family,baldi2014sparse}. In a more recent paper, a general technique was proposed to form the generator matrix of a QC-LDPC code by computing the permanents of some sub-matrices of the parity-check matrix with appropriate modifications based on rank deficiencies \cite{batgen}.

The sparse parity-check matrix is often represented as an array of circulant permutations, which facilitate efficient implementation but will typically have a number of linearly dependent rows. Although Gaussian elimination can be employed to compute the rank with a complexity of $\mathcal{O}(n^3)$, it is desirable both from the perspective of computational complexity and from the perspective of yielding insight about the code structure to have an analytic way to compute the rank, particularly for classes of algebraic QC-GLDPC codes. Methods to compute the rank of QC-LDPC codes have been investigated, including approaches involving Fourier transforms \cite{zhl+10} and the matrix polynomial representation \cite{ywc18}. However, these approaches are limited to certain code parameters.  In \cite{sgm22}, we presented a method to compute the rank of any parity-check matrix representing a QC-LDPC code, and hence the dimension of the code, by using the minors of the corresponding polynomial
parity-check matrix. %This formula, %recalled in Section~\ref{sec:background},  
%can be applied   to compute the rank of the QC-GLDPC codes as well.  

In this paper,  we first summarize and extend our results for QC-LDPC codes from \cite{sgm22}, then show how the approach extends naturally to QC-GLDPC codes. Using the QC-GLDPC constructions from \cite{lom18b} and \cite{lrc08} as examples, we demonstrate how to compute polynomial generator matrices, allowing efficient encoding. It is shown that the method also applies to mixed constructions, where the parity-check matrix has a mixture of generalized and single parity-check constraints. We consider the design of these codes and show that they can have large minimum distance. In particular, the generator matrices will often have rows that are equal (or close to) the minimum distance of the code, giving tight upper bounds that are hard to obtain otherwise. We discuss how the approach can also be used to determine the rank of the parity-check matrix of QC-GLDPC codes. Throughout the paper, we show that by applying double graph-liftings, the code parameters can be improved without affecting the ability to obtain a polynomial generator matrix. Importantly, this allows the code designer to finely tune the number of generalized constraints to optimize the performance/rate trade-off. Computer simulation results confirm the good performance promised by the constructed QC-GLDPC codes with high girth and large minimum distance.

The paper is organized as follows. Section \ref{sec:background} first provides some necessary background regarding QC-LDPC and QC-GLDPC codes. Sections \ref{sec:rank} and \ref{sec:genLDPC} establish theory regarding the rank and construction of polynomial generator matrices for QC-LDPC codes. Following this, we extend the results to QC-GLDPC codes in Section \ref{sec:genGLDPC}. The paper is concluded in Section \ref{sec:conc}.

\section{Definitions, Notation, and Background}\label{sec:background}
%
%\subsection{Notations and basic definitions}
We use the following notation. For any positive integer $L$,
$[L]$ denotes the set $\{  1,2, \ldots, L \}$. 
%By $\Ftwo^n$ and $\Ftwo^{m \times n}$ we will mean, respectively, a row vector
%over $\Ftwo$ of length $n$ and a matrix over $\Ftwo$ of size $m \times n$,
%with a similar meaning given to $\shortFtwoxmodN^n$, $\shortFtwoxmodN^{m
%  \times n}$. In the
%following we will use the convention that indices of vector entries start at
%$0$ (and not at $1$), with a similar convention for row and column indices of
%matrix entries.
%
For any $n_c\times n_v$ matrix $\matr{M}$, we let $\matr{M}_{\set{I}, \set{J}}$ be the
sub-matrix of $\matr{M}$ that contains only the rows of $\matr{M}$ whose index
appears in the set $\set{I}$ and only the columns of $\matr{M}$ whose index
appears in the set $\set{J}$; if $\set{I}$ equals the set of all row indices
of $\matr{M}$, {\ie, $\set{I}=[n_c]$}, we will simply write $\matr{M}_{\set{J}}$. We use the
short-hand $\matr{M}_{\set{J} \setminus i}$ for $\matr{M}_{\set{J} \setminus
  \{ i \}}$. If $\set{I}$ and $\set{J}$ have the same cardinality,  we use  {$\Delta_{\set{I},\set{J}}=\det(M_{\set{I},\set{J}})$, and $\Delta_{\set{J}}=\det(M_{[n_c], \set{J}}).$}
  If the matrix $M$ is an $n_c\times n_v$ block matrix with blocks of size $N\times N$, we use the same notations $\matr{M}_{\set{I}, \set{J}}$, $\matr{M}_{\set{J}}$, $\matr{M}_{\set{J} \setminus i}$, $\Delta_{\set{I},\set{J}}$  and $\Delta_{\set{J}}$ with the corresponding rows and columns considered when defining them  to be $N \times N$ block rows and columns. 
%For any matrix $\matr{M}$, we let $\matr{M}_{\set{R}, \set{S}}$ be the
%sub-matrix of $\matr{M}$ that contains only the rows of $\matr{M}$ whose index
%appears in the set $\set{R}$ and only the columns of $\matr{M}$ whose index
%appears in the set $\set{S}$. If $\set{R}$ equals the set of all row indices
%of $\matr{M}$, we will omit in $\matr{M}_{\set{R}, \set{S}}$ the set $\set{R}$
%and we will simply write $\matr{M}_{\set{S}}$.  Moreover, we will use the
%short-hand $\matr{M}_{\set{S} \setminus i}$ for $\matr{M}_{\set{S} \setminus
%  \{ i \}}$.
%

An LDPC code $\cC$ %of length $n$, dimension $k$, and minimum Hamming distance $\dmins$ 
is described as the nullspace of a %n $(n-k) \times n$ (scalar) \
parity-check matrix $\matr{H}$ % \in \GF{2}^{(n-k) \times n}$, i.e., $\code{C} = \bigl\{ \vc \in
%\Ftwo^n \ \bigl| \ \matr{H} \cdot \vc^\tr = \vect{0}^\tr \bigr. \bigr\}$,
 to which we associate a Tanner
graph~\cite{tan81} in the usual way. The girth, denoted $\girth(\matr{H})$, of a graph is the length of the shortest cycle in the graph. 
%Vectors are column vectors rather than row vectors. 

\subsection{Protographs and Polynomial Representations}
A protograph \cite{tho03,ddja09} is a small  bipartite graph  represented by an $n_c\times n_v$ parity-check or \emph{base} biadjacency matrix  $B$ with non-negative integer entries $b_{ij}$. The parity-check matrix $H$ of a protograph-based LDPC block code can be created by replacing each non-zero entry $b_{ij}$  by a sum of $ b_{ij}$ non-overlapping $N\times N$ permutation matrices and a zero entry by the $N\times N$ all-zero matrix. Graphically, this operation is equivalent to taking an $N$-fold graph cover, or ``lifting'', of the protograph. We denote the $N\times N$ circulant permutation matrix where the entries of the $N\times N$ identity matrix $I$ are shifted to the left by $r$ positions modulo $N$ as $I_r$. 

A QC-LDPC code of length $n = \horsize N$ is a protograph-based LDPC code, for which the $N\times N$ lifting permutation matrices are all circulant matrices  $I_r$.  
Thus a QC  code has an ${\vertsize N \times \horsize N}$ parity-check matrix $\matrH$ of
the form 
\begin{align}\label{QC} 
  \matrH
   = \begin{bmatrix}
         \matrH_{1,1}   & \matrH_{1,2}  & \cdots & \matrH_{1,\horsize} \\
         %\matrH_{2,1}   & \matrH_{2,2}  & \cdots & \matrH_{2,\horsize} \\
         \vdots          & \vdots          & \ddots & \vdots \\
         \matrH_{\vertsize,1} & \matrH_{\vertsize,2} & \cdots & 
           \matrH_{\vertsize,\horsize}
       \end{bmatrix},
       % \in \Ftwo^{\vertsize N \times \horsize N},
       \end{align}
where the ${N \times N}$ sub-matrices  $\matrH_{i,j}$ are circulant;  applying equal circular shifts to
each length $N$ sub-block of a codeword results in a codeword.  
With the help of the well-known isomorphism between the ring of circulant
matrices over the binary field $\Ftwo$ and the ring {$\Ftwo[x]/(x^N-1)$} of $\Ftwo$-polynomials
modulo $x^N - 1$ (see, \eg, \cite{Lally2000}), a QC-LDPC code can
also be described by an $\vertsize \times \horsize$ polynomial parity-check matrix
over $\Ftwo[x]/(x^N-1)$. This polynomial description of 
QC-LDPC codes makes them attractive in practice, because they can be
encoded efficiently using approaches like in~\cite{lcz+06}
and decoded efficiently using belief-propagation-based decoding
algorithms. 

In particular, with the ${\vertsize N \times \horsize N}$ parity-check matrix $\matrH $ %\in \Ftwo^{\vertsize N \times \horsize N}$ 
described above  we associate the polynomial parity-check matrix $\matrH(x) \in
\shortFtwoxmodN^{\vertsize \times \horsize}$ as 
%\vspace{-0mm}
\begin{align} 
  \matrH(x) \label{QC2}
    = \begin{bmatrix}
         h_{1,1}(x)   & h_{1,2}(x)  & \cdots & h_{1,\horsize}(x) \\
         %h_{2,1}(x)   & h_{2,2}(x)  & \cdots & h_{2,\horsize}(x) \\
         \vdots      & \vdots      & \ddots & \vdots \\
         h_{\vertsize,1}(x) & h_{\vertsize,2}(x) & \cdots & 
           h_{\vertsize,\horsize}(x)
       \end{bmatrix}.
\end{align}
%where $h_{i,j}(x)\in \Ftwo[x]/(x^N-1)$.  %\defeq \sum_{s=0}^{N-1} h_{j,i,s,0} x^s$. 
Here $h_{i,j}(x)=x^r$ corresponds to the circulant permutation matrix $I_r$. Moreover, with any vector  $$\vc = (c_{1,0}, \ldots, c_{1,N-1}, \ldots, c_{\horsize,0}, \ldots,
c_{\horsize,N-1}) $$ in $\Ftwo^{\horsize N}$,  we associate the polynomial vector 
%\begin{align*}
$  \vc(x)
    = \big(
         c_1(x),  \ldots, c_{\horsize}(x)
       \big)$
      % \in \shortFtwoxmodN^n,$
%\end{align*}
where $c_i(x) \defeq \sum\limits _{s=0}^{N-1} c_{i,s} x^s$. Then, the condition 
%\begin{align*}
$  \matrH \cdot \vc^\tr
    = \vect{0}^\tr $ 
          (in $\Ftwo$)
%\end{align*}
is equivalent to 
%\begin{align*}
 $ \matrH(x) \cdot \vc(x)^\tr
    = \vect{0}^\tr $ 
     in  $\Ftwo[x]/(x^N-1).$
 For the remainder of this paper, $H$ will be an $n_cN\times n_vN$ matrix over $\Ftwo$ and $H(x)$ will be  the corresponding $n_c\times n_v$ {polynomial} matrix.  % over $\Ftwo[x]/(x^N+1)$.

\subsection{GLDPC Codes}\label{GLDPC}
Let a \emph{component code} corresponding to one of the $Nn_c$ constraints, $c_i$, be represented by a $p_i\times q_i$ parity-check matrix (where a conventional single parity-check (SPC) constraint corresponds to the $1\times q_i$ all-ones matrix). The constraint matrix of a QC-GLDPC code is also represented as \eqref{QC} and \eqref{QC2}, however the full parity-check matrix is obtained by replacing each one entry with the corresponding column of the component parity-check matrix and zero entry with an all-zero column vector of the same dimension. Consequently, the design rate of the GLDPC code is %$R = 1-\frac{\sum\limits_{i=1}^{n_c}m^{c_i}}{n_v}$
\begin{equation}
R = 1-\frac{\sum_{i=1}^{n_c}{p_i}}{n_v}.
\end{equation}
Conventionally, BP decoding is performed on the constraint graph, corresponding to \eqref{QC}, where component decoding can be performed as desired, e.g., BCJR, BP, etc., with different performance/complexity trade-offs. This paper focuses on the design and efficient encoding of GLDPC codes.

%%%%%%%%%%%%%%%%%%%%%%%%%%
\section{The Rank of QC Matrices} \label{sec:rank}
%%%%%%%%%%%%%%%%%%%%%%%%%
This section gives a simple formula for the rank of $H$ and hence, the dimension of its associated code,  that is a function of the full minors of the polynomial matrix $H(x)$. We note the paper ~\cite{ywc18}  in which the authors had some particular cases based on the attempt to triangularize the matrix $H$. They do this only for the case  $n_c=3$.  Our formula holds for any general matrix of size $n_cN\times n_vN$. In \cite{sgm22}, we also gave a general triangular form into which any QC matrix can be written using elementary row operations and column permutations. Since these operations  leave the dimension and the weight enumerator of the code unchanged, we can obtain an alternative way to compute the dimension of the code and thus completely solve the task of  the paper~\cite{ywc18}.% done there only for $n_c=3$. 

The following theorem gives the formula for the rank {over $\Ftwo$} of a {scalar} matrix using the computation of the degree of the minors of the {corresponding polynomial matrix}.  

\begin{theorem}  \label{rank}  We define $\gamma_0\defeq 1$, and, for all $i\in[n_c]$, we  define the following  in $\Ftwo[x]$: %\vspace{-.1in}
\begin{align*}&\gamma_i(x)\defeq \gcd\{ \Delta_{\set{I},\set{J}}(H(x))\mid  |\set{I}|=|\set{J}|=i\}, 
%&\gamma_i(x)\defeq \gcd\{\rm \Delta_i}(H(x)\}, \text{ in }\Ftwo[x], i\in [n_c], \gamma_0\defeq 0, \\%\text{ in }  \Ftwo[x]\\
%d_i\defeq &\gamma_i/\gamma_{i-1},\\
\quad d_i(x) \defeq \gcd (\gamma_i(x)/\gamma_{i-1}(x), x^N+1), \end{align*}   where, by assumption, 
$\gcd(0/0,x^N+1)\defeq \gcd(0,x^N+1)=x^N+1.$   %\begin{align*}
%&\set{S} \defeq [n_c], \text{ w.l.o.g., }
%S_i\defeq S\cup \{n_c+i\}, 1\leq i\leq n_v-n_c,\\   
%&\Delta_{S_i \setminus j} \defeq \det\big( \matrH_{\set{S}_i \setminus j}(x) \big), 1\leq j\leq n_c. %d''_i(x) \defeq& \gamma_{n_c}/d_i,
%f_i\defeq &\frac{x^N+1}{d'_i}.
%\\\vspace{-.29in}%\end{align*} 
%\begin{align*}
Then, $$\rank_{\Ftwo}(H)= %\sum\limits_{i=1}^{n_c} (N-\deg d_i(x)) =  
n_c\cdot N -\sum\limits_{i=1}^{n_c} \deg d_i(x) \Rightarrow 
\dim(\cC)=(n_v-n_c)\cdot N +\sum_{i=1}^{n_c} \deg d_i(x).  $$ %\end{align*}
\end{theorem}

\begin{IEEEproof} In $\Ftwo[x]$, $H(x)$ is equivalent, after elementary row and column  operations that leave its rank (but not its minimum distance) invariant, to its Smith normal form 
$$\Gamma(x) \defeq \left[\begin{array}{c|c} \diag( \frac{\gamma_1}{\gamma_0}(x), \frac{\gamma_2}{\gamma_1}(x), \ldots , \frac{\gamma_{n_c}}{\gamma_{n_c-1}}(x)) & 0_{n_c\times (n_v-n_c)} \end{array} \right].$$  Equivalently, there exist invertible matrices $U(x), V(x)$ such that $U(x)H(x)V(x)=\Gamma(x) $ in $\Ftwo[x]$ and, therefore, such that 
$U(x)H(x)V(x)=\Gamma(x) \mod x^N-1$. The rank of the binary $n_cN\times n_vN$ matrix $H$ is thus equal to that of the binary $n_cN\times n_vN$ matrix $\Gamma$ corresponding to the $n_c\times n_v$ polynomial matrix $\Gamma(x)$,  which is equal to the sum of the ranks of the $N\times N$ circulant  matrices corresponding to the polynomials $\gamma_i(x)/\gamma_{i-1}(x)$. Since each individual  rank is $N- \deg (\gcd (\gamma_i(x)/\gamma_{i-1}(x), x^N+1)) \defeq N-\deg d_i(x)$, we obtain  that 
$$\rank_{\Ftwo}(H)= \rank_{\Ftwo}(\Gamma)=\sum\limits_{i=1}^{n_c} (N-\deg d_i(x)) = n_c\cdot N -\sum\limits_{i=1}^{n_c} \deg d_i(x) $$
and, equivalently, %\Rightarrow 
$\dim(\cC)=(n_v-n_c)\cdot N +\sum_{i=1}^{n_c} \deg d_i(x)$.  %  each being $N-\deg d_i(x). $ 
\end{IEEEproof} 
       %
%We also recall the  theorem that computes the rank of any matrix $H$ based on the degree of the polynomial minors of the associated $H(x)$.
%\begin{theorem}  [~\cite{sgm22}]\label{generalcase} Let { $H$ be a $n_cN\times n_vN$ matrix over $\Ftwo$ and}  $H(x)$ be {the corresponding}  $n_c\times n_v$ {polynomial} matrix.  % over $\Ftwo[x]/(x^N+1)$.
%For all $i\in[n_c]$, we  define the following  in $\Ftwo[x]$: 
%\vspace{-5pt}
%\begin{align*}&\gamma_i(x)\defeq \gcd\{ \Delta_{\set{I},\set{J}}(H(x))\mid  |\set{I}|=|\set{J}|=i\}, \gamma_0\defeq 1, \\
%%&\gamma_i(x)\defeq \gcd\{\rm \Delta_i}(H(x)\}, \text{ in }\Ftwo[x], i\in [n_c], \gamma_0\defeq 0, \\%\text{ in }  \Ftwo[x]\\
%%d_i\defeq &\gamma_i/\gamma_{i-1},\\
%&d_i(x) \defeq \gcd (\gamma_i/\gamma_{i-1}, x^N+1), \text{  where} \\& \gcd(0/0,x^N+1)\defeq \gcd(0,x^N+1)=x^N+1. \text { Then, }\end{align*}\vspace{-8mm}
%\begin{align*}
%%&\set{S} \defeq [n_c], \text{ w.l.o.g., }
%%S_i\defeq S\cup \{n_c+i\}, 1\leq i\leq n_v-n_c,\\   
%%&\Delta_{S_i \setminus j} \defeq \det\big( \matrH_{\set{S}_i \setminus j}(x) \big), 1\leq j\leq n_c. %d''_i(x) \defeq& \gamma_{n_c}/d_i,
%%f_i\defeq &\frac{x^N+1}{d'_i}.
%%\\\vspace{-.29in}%\end{align*} 
%%\begin{align*}
%&\rank_{\Ftwo}(H)= %\sum\limits_{i=1}^{n_c} (N-\deg d_i(x)) =  
%n_c\cdot N -\sum\limits_{i=1}^{n_c} \deg d_i(x) \Rightarrow \end{align*}\vspace{-7mm}
%\begin{align*}
%&\dim(\cC)=(n_v-n_c)\cdot N +\sum_{i=1}^{n_c} \deg d_i(x).\end{align*}\vspace{-.17in}
%\end{theorem}\vspace{-5pt}
%CAN WE ADD SOME BRIEF BACKGROUND FROM ISIT TO SUPPORT THE EXAMPLES? WE use $f$s in the examples, but it is not clear how.

\begin{example} \label{binomial}
Let $$H(x)\defeq  \begin{bmatrix}   1+x^2 &1+x^4& 1+x^6& 1+x^8& 1+x^{16}\\
                            x^4+x^{12}& x^{20} +x^{22}& x^{30}+ x^{42}& x^{40} +x^{14}& 1+ x^{50}\\
                             1+x^4 &x^{30}+x^{24}& x^{12}+x^{14}& x^3+x^{13}& x+x^9  \end{bmatrix} $$         
be a $3\times 5$ polynomial matrix in $\Ftwo[x]$. Depending on $N$, its Tanner graph can have girth $6$. We will keep $N$ variable, and compute the rank of the $3N\times 5N$ matrix  $H$ obtained from $H(x)$ in the ring $\Ftwo[x]/(x^N+1)$. 
Using the Magma program and its command for computing the needed determinants,  $\gamma_i\defeq {\rm GCD}({\rm Minors}(H,i))$,  in $\Ftwo[x]$,
we obtain:
$$ 
\gamma_1=x^2+1, \quad \gamma_2=x^4+1, \quad \gamma_3=x^6 + x^4 + x^2 + 1, \quad 
\frac{\gamma_1}{\gamma_0}=\frac{\gamma_2}{\gamma_1}=\frac{\gamma_3}{\gamma_2}= x^2+1, $$ from which, for all  $i \in [3]$,
$d_i=\gcd(x^2+1, x^N+1)=\left\{\begin{matrix} x^2+1, & \text{ if }  N \text{ even,} \\
                             %\rank(H) =3(N-\deg(x^2+1))= 3(N-2)=3N-6\\
 x+1,& \text{ if }  N \text{ odd,}
%\rank(H) =3(N-\deg(x+1))=3(N-1)=3N-3
\end{matrix}\right.$ ~ and so  %\end{align*}
\begin{align*}
& d_i=\gcd(x^2+1, x^N+1)=\left\{\begin{matrix} x^2+1, & \text{ if }  N \text{ even,} \\
                           % \rank(H) =3(N-\deg(x^2+1))= 3(N-2)=3N-6\\
 x+1,& \text{ if }  N \text{ odd,}
%\rank(H) =3(N-\deg(x+1))=3(N-1)=3N-3
\end{matrix}\right.\Longrightarrow \\\end{align*}
the Smith Normal form of $H$, over $\Ftwo[x]$, is 
$$\begin{bmatrix} x^2 + 1 & 0&0&0&0\\       0& x^2 + 1  &    0   &    0    &   0\\    0&       0 &x^2 + 1 &      0  &     0\end{bmatrix}. $$
Therefore, the  formula for the rank of $H$ depends on the parity of $N$: $$\rank(H) = 3(N-\deg(d_i))=\left\{\begin{matrix} 3N-6, & \text{ if }  N \text{ even,}\\ 
3N-3,& \text{ if }  N \text{ odd}\end{matrix}\right..$$

We observe that  the same matrix gives a code of slightly  larger dimension if $N$ is even. For example, $N=45$, gives $H$ of rank 132 and the code of dimension 93, while $N=46$ gives a matrix of the same rank, so a slightly longer code with dimension 98, while $N=44$ gives $H$ of rank 126 and a slightly shorter code of dimension 94.  
\end{example}

%%%%%%%%%%%%%%%%%%%%%%%%%%%%%%%%%%%%%%%%%%%%%%%%%%%%%%%
\section{Generator Matrices for QC-LDPC Codes}\label{sec:genLDPC}
%%%%%%%%%%%%%%%%%%%%%%%%%%%%%%%%%%%%%%%%%%%%%%%%%%%%%%
In~\cite{sgm22} we gave a method (based on polynomial minors of $H(x)$) to construct polynomial generator matrices for QC-LDPC codes. We extend these results and show how to obtain sets of codewords that can be used to create generator matrices for QC codes. % to find polynomial generator matrices for QC-GLDPC codes. 
We extend here  the simple technique described in~\cite{Smarandache_2012} to construct codewords of codes described by
polynomial parity-check matrices.\footnote{This is, in turn, an extension of a  codeword construction
technique by MacKay and Davey~\cite[Theorem~2]{md01}.} %\footnote{Note that
%the paper~\cite{MacKay:Davey:01:1} deals with codes that are described by
%scalar parity-check matrices composed of commuting permutation sub-matrices,
%of which parity-check matrices composed of cyclically shifted identity
%matrices are a special case.}
The following two lemmas allow us to construct codewords and generator matrices, and thus, to give upper bound on the minimum distance of the code.  

\begin{lemma}
  \label{lemma:codeword:construction:1}

    Let $\set{S} = \{  i_1, i_2, \ldots, i_{\vertsize+1} \}$
  be an arbitrary size $(\vertsize {+}1)$ subset of $[\horsize]$. Let  $\vect{c}_{\set{S}}(x)$ be the vector defined as 
  %{\begin{align*}&
  %\framebox{ 
$$\vect{c}_{\set{S}}(x) = \left( c_{\set{S},1}(x), \ldots, c_{\set{S},\horsize}(x) \right)= $$
 $$ \begin{array} {cccccccc}  \big(0\cdots 0& \Delta_{\set{S} \setminus {i_1} }^\tr    & 0\cdots 0 &\Delta^\tr_{\set{S} \setminus {i_{2}} }  &0\cdots 0&  \cdots & \Delta^\tr_{\set{S} \setminus {i_{n_c+1}} }  &0\cdots 0\big)  \\&\uparrow&& \uparrow&&& \uparrow &\\&i_1&& i_2&&& i_{\vertsize +1} &\end{array}$$ 
%\end{align*}}
%OLD DAVID  
% {\small  $$\begin{matrix} \vect{c}_{\set{S}}(x) = ( c_{\set{S},1}(x), \ldots, c_{\set{S},\horsize}(x) \big)= ~\end{matrix}
%  \begin{array} {cccccccccccc}  \big(0&\cdots &0& \Delta_{\set{S} \setminus {i_1} }^\tr  & \cdots  & \Delta^\tr_{\set{S} \setminus {i_{2}} }  &\cdots &   \Delta^\tr_{\set{S} \setminus {i_{n_c+1}} }  &0&\cdots &0\big)\end{array}$$}% \\&&&i_1&& i_2&& i_{\vertsize +1} &&&&\end{array}$$} %\end{align*}}
i.e.,   
   %%\in  \left(\Ftwo[x]/(x^N-1)\right)^\horsize$ 
  %be a length-$\horsize$ vector defined
  %by %\footnote{Because the ring $\shortFtwoxmodr$ has characteristic $2$, we
%    could equally well define $c_i(x) \defeq \det\big( \matrH_{\set{S}
%      \setminus i}(x) \big)$ if $i \in \set{S}$.}
%  \begin{align*}
   $ c_{\set{S},i}(x)
      \defeq
        % \begin{cases} 
           \det^\tr\big( \matrH_{\set{S} \setminus i}(x)\big)  =\Delta_{\set{S} \setminus i}$  
             if $i \in \set{S}$  and  $ c_{\set{S},i}(x) \defeq 0$                                      
            otherwise.
      %  \end{cases}.
  %\end{align*}
  Then $\vect{c}_{\set{S}}(x)$ and $\frac{1}{a(x)}\vect{c}_{\set{S}}(x) $ are codewords in $\code{C}$, for every $a(x)$ a divisor of $\Delta_{\set{S} \setminus i} $ in $\Ftwo[x]$,   for all $i \in \set{S}$, where the vectors (and the divisions) are first computed over $\Ftwo[x]$ and then projected onto $\Ftwo[x]/(x^N+1)$.   
  
The vectors $\vect{c}_{\set{S}}(D)$ and $\frac{1}{a(D)}\vect{c}_{\set{S}}(D) $ are also codewords in the time-invariant convolutional code generated by $H(D)=H(x)\big|_{x=D}$.
 %  \footnote{The determinant of an $m \times m$-polynomial matrix $\matr{B} = [b_{j,i}(x)]_{j,i}$
%over $\Ftwo[x]$ is 
%%\begin{align*}
%$  \det(\matr{B})
%    = \sum_{\sigma}
%         \prod_{j \in [m]}
%           b_{j,\sigma(j)}(x),$
%%\end{align*}
%where the summation is over all $m!$ permutations of the set $[m]$. Then $\det^\tr$ transposes the polynomials, i.e., the exponents are taken with the negative sign modulo $N$.} 
%  An analogous construction yields codewords of the convolutional code
%  $\codeCconv$ defined by the polynomial parity-check matrix $\matrHconv(y) \in
%  \Ftwoylauser^{\vertsize \times \horsize}$.
\end{lemma}
%%%%%%%%%%%%%%%%%%%%%%%%%%%%%%%%%%%%
\begin{IEEEproof}
  Let $\vs^\tr(x) =
  \matrH(x) \cdot \vc_{\set{S}}^\tr(x)$ be the syndrome. Then, for any $j \in [\vertsize]$, 
    \begin{align*}
   s_{j}(x) 
%      = \sum_{i \in [\horsize]}
%           h_{j,i}(x) c_{\set{S},i}(x)
      = \sum_{i \in \set{S}}
           h_{j,i}(x)
           \cdot
           \det\big( \matrH_{\set{S} \setminus i}(x) \big)=   \left|
    \begin{array}{cccc}
      h_{j,i_1}(x)   & h_{j,i_2}(x)   & \cdots & h_{j,i_\vertsize +1 }(x) \\ 
      \hline
      h_{1,i_1}(x)   & h_{1,i_2}(x)   & \cdots & h_{1,i_\vertsize+1}(x) \\   \vdots        & \vdots        & & \vdots \\ 
%      h_{j,i_1}(x)   & h_{j,i_2}(x)   & \cdots & h_{j,i_\vertsize +1 }(x) \\
%      \vdots        & \vdots        &  & \vdots \\ 
      h_{\vertsize, i_1}(x) & h_{\vertsize,i_2}(x) & \cdots &  h_{\vertsize,i_\vertsize+1 }(x) \\ 
    \end{array}
    \right|=0, 
%    = \sum_{i \in \set{S}}
%         h_{j,i}(x)
%           \cdot
%          \det\big( \matrH_{\set{S} \setminus i}(x) \big).
 \end{align*}
since
  $s_{j}(x)$ is the co-factor expansion of the determinant of a $|\set{S}|
  \times |\set{S}|$ matrix which has determinant zero over $\Ftwo[x]$. 
  
  Since $a(x)$ is a divisor of all of the components, we can divide the codeword $\vc_{\set{S}}(x)$ by $a(x)$. Its syndrome will be the syndrome of $\vc_{\set{S}}(x)$ divided by $a(x)$, which is still zero in $\Ftwo[x]$. Therefore, $\vc_{\set{S}}(x)$ and $\frac{1}{a(x)}\vc_{\set{S}}(x)$  are  codewords in the associated convolutional code,  and therefore, when projected onto  $\Ftwo[x]/(x^N+1),$ they are 
also codewords in the QC code.    

Finally, since $\matrH(x) \cdot \vc_{\set{S}}^\tr(x)=0$  in $\Ftwo[x]$ and the division by $a(x)$ is performed in $\Ftwo[x]$,  we obtain that   $\matrH(D) \cdot \vc_{\set{S}}^\tr(D)=0$ and, therefore, the vectors $\vect{c}_{\set{S}}(D)$ and $\frac{1}{a(D)}\vect{c}_{\set{S}}(D) $ are also codewords in the time-invariant convolutional code generated by $H(D)=H(x)\big|_{x=D}$.
% \begin{align*}
%    \left|
%    \begin{array}{cccc}
%      h_{j,i_1}(x)   & h_{j,i_2}(x)   & \cdots & h_{j,i_\vertsize +1 }(x) \\ 
%      \hline
%      h_{1,i_1}(x)   & h_{1,i_2}(x)   & \cdots & h_{1,i_\vertsize+1}(x) \\   \vdots        & \vdots        & & \vdots \\ 
%      h_{j,i_1}(x)   & h_{j,i_2}(x)   & \cdots & h_{j,i_\vertsize +1 }(x) \\
%      \vdots        & \vdots        &  & \vdots \\ 
%      h_{\vertsize, i_1}(x) & h_{\vertsize,i_2}(x) & \cdots &  h_{\vertsize,i_\vertsize+1 }(x) \\ 
%    \end{array}
%    \right|.  
%  \end{align*} 
%  we obtain that $\vs(x) = \vect{0}$ and,  thus, $\vc(x)$
  %is  a codeword. %  in $\code{C}$.
%  Because $\Ftwoylauser$ is a field, and therefore a commutative ring, the
%  same argument holds also for a code like $\codeCconv$ that is defined by
%  a parity-check matrix over $\Ftwoylauser$.
\end{IEEEproof}

\begin{lemma}
  \label{lemma:codeword:construction:2}
 Let $s\in\{0,\ldots, \vertsize -1\}$, and let $\set{S} = \{  i_1, i_2, \ldots, i_{s+1} \}$
  be an arbitrary size $(s +1)$ subset of $[\horsize]$. Let $\set{T} = \{  j_1, j_2, \ldots, j_{s} \}$
  be an arbitrary size $s$ subset of $[\vertsize]$ (where a size 0 set is an empty set),   and let
    $$ \vect{c}_{\set{T}, \set{S}} = ( c_{\set{T}, \set{S},1}(x), \ldots, c_{\set{T}, \set{S},\horsize}(x) \big)=$$ $$  \begin{array} {cccccccccccc}  \big(0\cdots 0& (f(x) \Delta_{\set{T},\set{S} \setminus {i_1} })^\tr  %& f(x)^\tr \Delta^\tr_{\set{T},\set{S} \setminus {i_{2}} }  &\ldots 
 &0\cdots 0 & \cdots &  (f(x) \Delta_{\set{T},\set{S} \setminus {i_{s+1}} })^\tr  &0\cdots 0\big) \\&\uparrow&&&\uparrow &&&&\\&i_1&&& i_{s +1} &&&&\end{array}$$

    %{\small $$\begin{matrix} \vect{c}_{\set{T}, \set{S}} = ( c_{\set{T}, \set{S},1}(x), \ldots, c_{\set{T}, \set{S},\horsize}(x) \big)= \\~\end{matrix}
 % \begin{array} {cccccccccccc}  \big(0&\ldots &0& (f(x) \Delta_{\set{T},\set{S} \setminus {i_1} })^\tr&\! \ldots\!  %& f(x)^\tr \Delta^\tr_{\set{T},\set{S} \setminus {i_{2}} }  &\ldots 
 % &  (f(x) \Delta_{\set{T},\set{S} \setminus {i_{s+1}} })^\tr  &0&\!\ldots \!&0\big) \\&&&i_1&& i_{s +1} &&&&\end{array}$$}
   %%\in  \left(\Ftwo[x]/(x^N-1)\right)^\horsize$ 
  i.e.,  %\footnote{Because the ring $\shortFtwoxmodr$ has characteristic $2$, we
%    could equally well define $c_i(x) \defeq \det\big( \matrH_{\set{S}
%      \setminus i}(x) \big)$ if $i \in \set{S}$.}
  %\begin{align*}
   $ c_{\set{T},\set{S},  i}(x)
     \defeq
        % \begin{cases} 
        ( f(x) \Delta_{\set{T},\set{S} \setminus i})^\tr$
              if $i \in \set{S}$ and 
          $c_{\set{T},\set{S},  i}(x)
     \defeq 0$                                     
             otherwise. 
        % \end{cases}.
 % \end{align*}
Then $\vect{c}_{\set{T},\set{S}}(x)$ is a codeword in $\code{C}$ if and only if $f(x)\cdot\Delta_{\set{T}\cup j,\set{S}}=0$,  for all $j\in [\vertsize]\setminus T$.
%  An analogous construction yields codewords of the convolutional code
%  $\codeCconv$ defined by the polynomial parity-check matrix $\matrHconv(y) \in
%  \Ftwoylauser^{\vertsize \times \horsize}$.
\end{lemma}
%%%%%%%%%%%%%%%%%%%%%%%%%%%%%%%%%%%%
\begin{IEEEproof}
  Let $\vs^\tr(x) =
  \matrH(x) \cdot \vc_{\set{T},\set{S}}^\tr(x)$ be the syndrome. Then, for any $j \in [\vertsize]$, 
    \begin{align*}
   s_j(x)     & = \sum_{i \in \set{S}}
           h_{j,i}(x)
         \cdot  f(x)\cdot \det\big( \matrH_{\set{T}, \set{S} \setminus i}(x) \big)=f(x)\left|
    \begin{array}{cccc}
      h_{j,i_1}(x)   & h_{j,i_2}(x)   & \cdots & h_{j,i_s+1 }(x) \\ 
      \hline
      h_{j_1,i_1}(x)   & h_{j_1,i_2}(x)   & \cdots & h_{j_1,i_{s+1}}(x) \\   \vdots        & \vdots        & & \vdots \\ 
%      h_{j_l,i_1}(x)   & h_{j_l,i_2}(x)   & \cdots & h_{j_l,i_{s +1} }(x) \\
%      \vdots        & \vdots        &  & \vdots \\ 
      h_{j_s, i_1}(x) & h_{j_s,i_2}(x) & \cdots &  h_{j_s,i_{s+1} }(x) \\ 
    \end{array}
    \right| \\&=f(x)\cdot\left\{ \begin{matrix} 0, & \text{ if } j\in \set{T}, \\\Delta_{\set{T}\cup j,\set{S}},& \text{ if } j\in [\vertsize]\setminus\set{T}\end{matrix}\right.=\left\{ \begin{matrix} 0, & \text{ if } j\in \set{T}, \\ f(x)\cdot\Delta_{\set{T}\cup j,\set{S}},& \text{ if } j\in [\vertsize]\setminus\set{T}.\end{matrix}\right.%    = \sum_{i \in \set{S}}
%         h_{j,i}(x)
%           \cdot
%          \det\big( \matrH_{\set{S} \setminus i}(x) \big).
 \end{align*}
Therefore, 
  we obtain that $\vs(x) = \vect{0}$ and,  thus, $\vc_{\set{T},\set{S}}(x)$
  is  a codeword if and only if $f(x)\cdot\Delta_{\set{T}\cup j,\set{S}}=0$ for all $j\in [\vertsize]\setminus T$. %  in $\code{C}$.
%  Because $\Ftwoylauser$ is a field, and therefore a commutative ring, the
%  same argument holds also for a code like $\codeCconv$ that is defined by
%  a parity-check matrix over $\Ftwoylauser$.
\end{IEEEproof}
We will use these two lemmas to construct codewords and generator matrices for QC codes. We will give an example first. 
\begin{example} \label{Example-codewords} Let $H_1(x)=\begin{bmatrix} 1+x&1+x^2&(1+x^3)(1+x)& 1+x^3\end{bmatrix}.  $  We apply Lemma~\ref{lemma:codeword:construction:1} and obtain the following  12 codewords with $a(x)=x+1$:
\begin{align*}\vc_{\{1,2\}}=(1+x^2, 1+x, 0,0), \quad & \frac{1}{x+1}\vc_{\{1,2\}}=(1+x, 1, 0,0),\\
\vc_{\{1,3\}}=((1+x^3)(1+x), 0,1+x,0), \quad &\frac{1}{x+1}\vc_{\{1,3\}}=(1+x^3,0, 1, 0),\\
\vc_{\{1,4\}}=(1+x^3, 0,0, 1+x), \quad &\frac{1}{x+1}\vc_{\{1,4\}}=(1+x+x^2,0, 0, 1),\\
\vc_{\{2,3\}}=(0, (1+x^3)(1+x), 1+x^2,0), \quad &\frac{1}{x^2+1}\vc_{\{2,3\}}=(0,1+x+x^2, 1,0),\\
\vc_{\{2,4\}}=(0, 1+x^3, 0,1+x^2), \quad &\frac{1}{x+1}\vc_{\{2,4\}}=(0, 1+x+x^2,0, 1+x),\\
\vc_{\{3,4\}}=(0,0, 1+x^3, (1+x^3)(1+x)), \quad &\frac{1}{x+1}\vc_{\{3,4\}}=(0,0,1,1+x).
\end{align*}
We apply Lemma~\ref{lemma:codeword:construction:2},  for $s=0$,  and obtain the following 4 codewords:
\begin{align*} (f_1(x),0,0,0), \quad & \text{ where } f_1(x)(1+x)=1+x^N, \text{  in } \Ftwo[x], \\(0, f_2(x),0,0), \quad & \text{ where } f_2(x)\cdot\gcd(1+x^2, 1+x^N)=1+x^N, \text{  in } \Ftwo[x],  \\
(0, 0, f_3(x),0), \quad & \text{ where } f_3(x)\cdot\gcd(1+x+x^3+x^4, 1+x^N)=1+x^N, \text{  in } \Ftwo[x],  \\(0, 0,0, f_4(x)), \quad & \text{ where } f_4(x)\cdot\gcd(1+x^3, 1+x^N)=1+x^N, \text{  in } \Ftwo[x] .
\end{align*}
%Since $ f_1(x)\cdot\gcd(1+x^2, 1+x^N)=1+x^N$, $ f_1(x)\cdot\gcd(1+x+x^3+x^4, 1+x^N)=1+x^N $ and $f_1(x)\cdot\gcd(1+x^3, 1+x^N)=1+x^N$,
We note that $(0, f_1(x),0,0),  (0,0, f_1(x),0), (0,0,0, f_1(x))$ are also codewords. 

Since $\rank(H)= N-\deg(1+x)=N-1$, the rank of $G$ needs to be $3N+1$. So in order to achieve the desired rank, we need to use some of the 12 codewords in the top list, together with one or more of the 4 codewords  from the bottom list, depending on $N$. (This will be elaborated on below after we introduce Theorem \ref{case1}.) 
 For example, 
 $$G_1(x)\defeq \begin{bmatrix} 1+x^2& 1+x& 0&0 \\
  (1+x^3)(1+x)& 0&1+x&0\\
1+x^3& 0&0& 1+x\\\hline
f_1(x)&0&0&0 \\
0& f_2(x)&0&0 \\
0& 0& f_3(x)&0  \\
0& 0&0& f_4(x)
\end{bmatrix} , G_2(x)\defeq\begin{bmatrix} 1+x& 1& 0&0 \\
  1+x^3& 0&1&0\\
1+x+x^2& 0&0& 1\\\hline
f_1(x)&0&0&0 \\
\end{bmatrix} $$
are two generator matrices for this code. 
In this case, the second matrix has lower weight rows, so it would seem to be the natural choice; however, it  is not always the case that the second type will be more attractive, since when we divide a polynomial by another we do not always obtain a lower weight. In fact, sometimes, the weight increases drastically, for example when we divide $1+x^N$ by $1+x$. Therefore, both matrices could be considered and compared. They give different sets of codewords, and thus upper bounds on the minimum distance of the code. In this case, we obtain $ d(\cC)\leq  3$ from the second matrix. \end{example} 

%\begin{corollary} The following sets of codewords 
%For all $ i\in [m]$, $m=n_v-n_c$, we apply Lemma~\ref{lemma:codeword:construction:1} to each of the sets $S_i= S\cup \{n_c+i\}$, 
%to obtain the linear independent set of codewords 
%\begin{align*} \vect{c}_{S_1} \defeq &\left(\begin{matrix}\Delta_{S_1\setminus 1})^\tr & (\Delta_{S_1\setminus 2})^\tr &
%  \cdots &(\Delta_{S_1\setminus n_c})^\tr & \Delta_{\set{S}}&0& \cdots &0\end{matrix}\right)\\ 
% \vect{c}_{S_2} \defeq &\left(\begin{matrix}\Delta_{S_2\setminus 1})^\tr & (\Delta_{S_2\setminus 2})^\tr &
%  \cdots &(\Delta_{S_2\setminus n_c})^\tr & 0&\Delta_{\set{S}}&\cdots &0\end{matrix}\right)\\
%\vdots& \\
%%  \vect{c}_{S_i} \defeq &\left(\begin{matrix}\Delta_{S_i\setminus 1})^\tr & (\Delta_{S_i\setminus 2})^\tr &
%%  \cdots &(\Delta_{S_i\setminus n_c})^\tr &0&\cdots &0 & \Delta_{\set{S}}&0&0\end{matrix}\right)\\\vdots &\\
%    \vect{c}_{S_m} \defeq &\left(\begin{matrix}\Delta_{S_m\setminus 1})^\tr & (\Delta_{S_m\setminus 2})^\tr &
%  \cdots &(\Delta_{S_m\setminus n_c})^\tr &0&0&\cdots &\Delta_{\set{S}}\end{matrix}\right),  \end{align*}
%  which form  rows of the generator matrix $G$. 
%\end{corollary} 

As we saw in  Example~\ref{Example-codewords},  the two lemmas can be used to obtain a variety of generator matrices, with the goal of getting one of the lowest weight, equal to the minimum distance.  

 \begin{theorem} \label{case1} 
 %If no $n_c\times n_c$ polynomial submatrix of $H(x)$ is invertible, then 
 Suppose there exists  a subset $\set{S}$ of size $\vertsize$ of $[\horsize]$  such that $\gcd(\Delta_{\set{S}}, x^N+1)=1$, i.e., $\Delta_{\set{S}}$ is invertible in $\Ftwo[x]/(x^N+1)$. For simplicity,  and w.l.o.g.,  we assume that $\set{S} =[n_c]$. For all $ i\in [m]$, $m=n_v-n_c$, let  $S_i= S\cup \{n_c+i\}$, and let 
$G$ be the $m\times \horsize$ matrix defined as  
 $$G\defeq \begin{bmatrix} \vect{c}_{S_1} \\ \vect{c}_{S_2} \\ \vdots\\ \vect{c}_{S_m} %\\ \vect{c}_{[n_c]\setminus 1, \set{S}} \\  \vect{c}_{[n_c]\setminus 2, S}  \\ \vdots\\  \vect{c}_{[n_c]\setminus n_c, S} 
 \end{bmatrix} 
  = \begin{bmatrix} \Delta_{S_1\setminus 1}^\tr &
  \cdots &\Delta_{S_1\setminus n_c}^\tr & \Delta^\tr_{\set{S}}&0& \cdots &0\\ 
\Delta_{S_2\setminus 1}^\tr  &
  \cdots &\Delta_{S_2\setminus n_c}^\tr & 0&\Delta^\tr_{\set{S}}&\cdots &0\\
\vdots&&\vdots&\vdots &\vdots&&\vdots \\
%  \vect{c}_{S_i} \defeq &\left(\begin{matrix}\Delta_{S_i\setminus 1})^\tr & (\Delta_{S_i\setminus 2})^\tr &
%  \cdots &(\Delta_{S_i\setminus n_c})^\tr &0&\cdots &0 & \Delta_{\set{S}}&0&0\end{matrix}\right)\\\vdots &\\
    \Delta_{S_m\setminus 1}^\tr  &
  \cdots &\Delta^\tr_{S_m\setminus n_c} &0&0&\cdots &\Delta^\tr_{\set{S}}\end{bmatrix}.$$
  %\\\hline
%    (\Delta_{S_1 \setminus 1})^\tr  &  \cdots &(\Delta_{S_1 \setminus n_c})^\tr \\ 
%                                                             \vdots &  & \vdots \\ 
%                                                             (\Delta_{S_{m}\setminus 1})^\tr  & \cdots&    (\Delta_{S_{m}\setminus n_c})^\tr \\\hline  
 
                           %          \begin{array}{c|c}  \begin{matrix}         f(x) \Delta_{[n_c]\setminus 1,\set{S} \setminus {1} })^\tr& \ldots 
%   %& f(x)^\tr \Delta^\tr_{\set{T},\set{S} \setminus {i_{2}} }  &\ldots 
%  &  (f(x) \Delta_{[n_c]\setminus 1,\set{S} \setminus {n_c}} )^\tr  \\\vdots&&\vdots\\
%(f(x) \Delta_{[n_c]\setminus n_c,\set{S} \setminus {1} })^\tr& \ldots  %& f(x)^\tr \Delta^\tr_{\set{T},\set{S} \setminus {i_{2}} }  &\ldots 
%  &  (f(x) \Delta_{[n_c]\setminus n_c,\set{S} \setminus {n_c}} )^\tr \end{matrix}&0  \end{array}
\noindent Then, $G$ and $(\Delta^\tr_{\set{S}})^{-1}G$ are generator matrices for $\cC$, where the second one is in  standard form.  %together with the rank of the top part equals the dimension of the code.  
 \end{theorem}

 %Then the $m\times \horsize$ matrix  below is a polynomial  generator matrix for $\cC$,
   %  {$$ \left[\begin{array}{c|c} \begin{matrix} (\Delta_{S_1 \setminus 1}\Delta_{\set{S}}^{-1} )^\tr  &  %(\Delta_{S_1 \setminus 2}\Delta_{\set{S}}^{-1})^\tr &
% \cdots &(\Delta_{S_1 \setminus n_c}\Delta_{\set{S}}^{-1})^\tr \\ % (\Delta_{S_2\setminus 1}\Delta_{\set{S}}^{-1})^\tr & %(\Delta_{S_2\setminus 2}\Delta_{\set{S}}^{-1})^\tr & 
% %\cdots &(\Delta_{S_2\setminus n_c}\Delta_{\set{S}}^{-1} )^\tr\\ 
% \vdots & %\vdots & 
% & \vdots \\ (\Delta_{S_{m}\setminus 1}\Delta_{\set{S}}^{-1})^\tr  & %(\Delta_{S_{m}\setminus 2}\Delta_{\set{S}}^{-1})^\tr&
% \cdots&    (\Delta_{S_{m}\setminus n_c}\Delta_{\set{S}}^{-1})^\tr \end{matrix}& \matr{I}_{m\times m}\end{array}\right], 
 %$$
%where $m\defeq n_v-n_c,$ $S_i= S\cup \{n_c+i\}$, for all $ i\in [m]$,  $\Delta_{S_i \setminus j}$ $ \defeq  \det\big( \matrH_{\set{S}_i \setminus j}(x) \big)$, for all $ j\in [n_c]$,  $\diag_{m}( \Delta_{\set{S}})$ is a diagonal $m\times m$ matrix with  each diagonal entry equal to $\Delta_{\set{S}}$.

\begin{IEEEproof}   For all $ i\in [m]$, $m=n_v-n_c$, we apply Lemma~\ref{lemma:codeword:construction:1} to each of the sets $S_i= S\cup \{n_c+i\}$, 
to obtain the linearly independent set $\{\vect{c}_{S_1}, \vect{c}_{S_2}, \cdots, \vect{c}_{S_m} \}$ of $n_v-n_c$ rows of $G$, of rank $mN=n_vN-n_cN=\dim(C)$, as needed. The second matrix has the same property.
\end{IEEEproof}

For brevity in examples, we will often drop the set notation in subscripts where there is no ambiguity and single digit integers. For example, if $S=\{1,2,3\}$, we write $\Delta_{123}=\det(H_{123})$.

%respectively, 
%$\diag_{n_c}( \Delta_{\set{S}})$ is a diagonal $n_c\times n_c$ matrix with  each diagonal entry equal to $\Delta_{\set{S}}$, and \\\\
\begin{example} (AR4JA codes.)  Let $H(x)$ be the polynomial parity-check matrix given by
\vspace{-1mm}
\begin{equation*}
H(x)=\left[\begin{matrix} 0&0&1&0&1+x \\ 1&1&0&1&x+x^2+x^3 \\ 1&x+x^2&0&1+x^3&1 \end{matrix}\right],
\end{equation*}
with $N=4$. The rows of the matrix $G(x)$ below  are linearly independent codewords for the code $\cC$ given by $H$. Since $\Delta_{123}=\det(H_{123})=x^2+x+1 $ is irreducible in $\Ftwo[x]/(x^4+1)$, the matrix $G(x)$ has  rank $2N=8$ and thus forms  a generator matrix for the code $\cC$, where
%\begin{equation*}
\vspace{-3mm} 

\begin{equation*}
G(x)=\left[\begin{matrix} \Delta_{234}^\tr&\Delta_{134}^\tr&\Delta_{124}^\tr&\Delta_{123}^\tr&0 \\ \Delta_{235}^\tr&\Delta_{135}^\tr&\Delta_{125}^\tr&0&\Delta_{123} ^\tr
%\\ \Delta_{245}&\Delta_{145}&0&\Delta_{125}&\Delta_{124} \\ \Delta_{345}&0&\Delta_{145}&\Delta_{135}&\Delta_{134} \\ 0&\Delta_{345}&\Delta_{245}&\Delta_{235}&\Delta_{234} 
\end{matrix}\right]=
\end{equation*}\vspace{-1mm}
\begin{equation*}
{\left[\begin{matrix} (x+1)^3&x&0&x^3+x^2+1&0 \\x^3+x^2+1&(x+1)^3&x+1&0&x^3+x^2+1
%\\ \Delta_{245}&\Delta_{145}&0&\Delta_{125}&\Delta_{124} \\ \Delta_{345}&0&\Delta_{145}&\Delta_{135}&\Delta_{134} \\ 0&\Delta_{345}&\Delta_{245}&\Delta_{235}&\Delta_{234} 
\end{matrix}\right].}
\end{equation*}
%\vspace{2mm}
 %\end{equation*}
 %where $\Delta_{123}=\det(S_{123})=x^3+x^2+1. $ 
% 
%
%where \vspace{-6mm}
%\begin{align*}
%\Delta_{123}&=\det(S_{123})=x^2+x+1 \\
%\Delta_{124}&=\det(S_{124})=0 \\
%\Delta_{125}&=\det(S_{125})=x^3+1 \\
%\Delta_{134}&=\det(S_{134})=x^3 \\
%\Delta_{135}&=\det(S_{135})=x^3+x^2+x+1 \\
%%\Delta_{145}&=\det(S_{145})=x^3+1 \\
%\Delta_{234}&=\det(S_{234})=x^3+x^2+x+1 \\
%\Delta_{235}&=\det(S_{235})=x^2+x+1 \\
%%\Delta_{245}&=\det(S_{245})=0 \\
%%\Delta_{345}&=\det(S_{345})=x^3
%\end{align*}
The vectors  $(\Delta^\tr_{245},\Delta^\tr_{145},0,\Delta^\tr_{125},\Delta^\tr_{124}),$ $  (\Delta^\tr_{345},0,\Delta^\tr_{145},\Delta^\tr_{135},\Delta^\tr_{134}), $ $ (0,\Delta^\tr_{345},\Delta^\tr_{245}, \Delta^\tr_{235},\Delta^\tr_{234}),$ are also codewords, where  
 $\Delta_{145}=\det(H_{145})=x+1,$ $\Delta_{245}=\det(H_{245})=0$,  and $\Delta_{345}=\det(H_{345})=x,$ 
  but they are linear combinations of the ones displayed in the matrix $G(x)$, so they are not needed for the generator matrix. Displaying them is useful, however, since  the nonzero  minimum of the weights of the five codewords  gives an upper bound on the minimum distance of the code. In this case, the codeword $ (\Delta^\tr_{245},\Delta^\tr_{145},0,\Delta^\tr_{125},\Delta^\tr_{124})= (0,x+1, 0, x+1,0) $ has weight 4, which is in fact the minimum distance of the code; this code has parameters $[20,8,4]$.  

Multiplying the two rows of $G$ by  $(x^3+x^2+1)^{-1} =x^2 + x + 1$ gives  a generator matrix in standard form, 
%\vspace{-3mm}
%\begin{multline*}\hspace{-4mm}
$$G(x)={\left[\begin{matrix} x^3 + x^2 + x + 1&x^3 + x^2 + x&0 &1 &0 \\ 1&x^3 + x^2 + x + 1 & x^3+1 &0&1 
\end{matrix}\right]}, $$%\end{multline*}}
while \vspace{-2mm}
%\vspace{-5mm}
$$\hspace{5mm}H(x)=\begin{bmatrix} 1&0&0 & x^3 + x^2 + x + 1&1  \\ 0&1&0&  x + x^2 + x^3 &x^3 + x^2 + x + 1\\ 0&0&1& 0 &x+1\end{bmatrix} $$
is a polynomial parity-check matrix  in systematic form.
\end{example}

 \begin{remark}\label{case2}
 In  Example~\ref{Example-codewords},  we saw how to extend Theorem~\ref{case1} using  Lemma~\ref{lemma:codeword:construction:2}  to obtain a generator matrix in the case $\gcd(\gamma_{n_c}(x), x^N+1)=g(x)\neq 1$, for $\gamma_{n_c}(x)= \gcd\{ \Delta_{\set{I},\set{J}}(H(x))\mid  |\set{I}|=|\set{J}|=n_c\}. $ We choose $f(x)$ such that $f(x)g(x)=x^N+1$ in $\Ftwo[x]$,  and apply Lemma~\ref{lemma:codeword:construction:2} to, for example,  the sets $[n_c]\setminus i, \set{S}$, to obtain extra vectors,   $\vc_{[n_c]\setminus i,\set{S}}(x)$,  that can be added to the matrix $G$, of rank $m(N-\deg g(x))$, to increase the rank to the dimension of $\cC$, which can be computed using Theorem~\ref{rank} as $\dim(\cC)=m\cdot N +\sum_{i=1}^{n_c} \deg d_i(x).  $    
 
 For example, if we divide each of the rows of $G$ by $\gamma_{n_c}(x)$ in $\Ftwo[x]$ and then add  the following codewords,  $$ \begin{bmatrix}(f(x) \Delta_{[n_c]\setminus 1,\set{S}  })^\tr& \ldots 
   %& f(x)^\tr \Delta^\tr_{\set{T},\set{S} \setminus {i_{2}} }  &\ldots 
  &  (f(x) \Delta_{[n_c]\setminus 1,\set{S} } )^\tr  &0&0&\cdots&0\\\vdots&&\vdots\\
(f(x) \Delta_{[n_c]\setminus n_c,\set{S} })^\tr& \ldots  %& f(x)^\tr \Delta^\tr_{\set{T},\set{S} \setminus {i_{2}} }  &\ldots 
  &  (f(x) \Delta_{[n_c]\setminus n_c,\set{S} } )^\tr    \diag_{m}(\Delta_{\set{S}})^\tr &0&0&\cdots&0\end{bmatrix}.  $$
  we obtain a generator matrix for the code. Without dividing, the top part of the matrix obtained from Theorem~\ref{case1} has a lower rank than the one when we divide, so we would  need to add more codewords obtained from Lemma~\ref{lemma:codeword:construction:2}, possibly all (as is the case in Example~\ref{Example-codewords}),  to increase the rank of the matrix to equal the dimension.  \end{remark}   
 
 In  Section~\ref{GLDPC}, we will show how to obtain generator matrices for QC-GLDPC codes  and we will encounter this situation  several times. In addition, we will need the theorem below, which simplifies our task, by allowing us to reduce finding a generator matrix for an $[n_vN,n_vN-\rank(H)]$ code to finding a generator matrix for a shorter code of the same dimension.

 \begin{theorem}\label{thm-short} Let $H_{short}\defeq H_{m_1\times n} $ be an $m_1\times n$ matrix and $A\defeq A_{m_2\times n} $ be an $m_2\times n$ matrix.  
Let $$H\defeq H_{(m_1+m_2)\times (n+m_2)}  \defeq\begin{bmatrix} H_{short} &{0}_{m_1\times m_2} \\A&{I}_{m_2\times m_2} \end{bmatrix}, $$   $G_{short}$ be  a generator matrix   for $\ker(H_{short})$, and  $g_{GC}\defeq  \begin{bmatrix} I_{n\times n}  & A^\tr\end{bmatrix}$ is a generator matrix for the code $\ker \begin{bmatrix} A&I_{m_2\times m_2}\end{bmatrix} %h_{GC}
$  in standard form. Then the matrix $$G\defeq \begin{bmatrix} G_{short} & G_{short}A^\tr\end{bmatrix}=G_{short}\begin{bmatrix} I_{n\times n}  & A^\tr\end{bmatrix}= G_{short}  \cdot g_{GC}$$ is a generator matrix for $\ker(H), $  where  the codewords  
   $(\vect{v}_1, \vect{v}_2) $, satisfy 
$$(\vect{v}_1, \vect{v}_2) = (\vect{v}_1, \vect{v}_1A^\tr)=\vect{v}_1 \begin{bmatrix} I_{n\times n}& A^\tr\end{bmatrix} =\vect{v}_1 \cdot g_{GC} ,$$ 
  and $$d_{min}(\ker(H))\geq d_{min}(\ker(H_{short})).$$ \end{theorem} 
\begin{IEEEproof} Let $(\vect{v}_1, \vect{v}_2)^\tr \in \ker(H)$, therefore,
$\vect{v}_1^\tr\in \ker(H_{short})$, and $\vect{v}_2 =\vect{v}_1A^\tr$. 
We have that  $$HG^\tr= \begin{bmatrix} H_{short} &\vect{0} \\A&I\end{bmatrix}\begin{bmatrix} G_{short}^\tr \\ (G_{short}A^\tr)^\tr\end{bmatrix}= \begin{bmatrix} H_{short}G_{short}^\tr\\ AG_{short}^\tr +(G_{short}A^\tr)^\tr
\end{bmatrix} = \begin{bmatrix} 0\\ 0\end{bmatrix},$$ since $H_{short}G_{short}^\tr=0.$ 
Further, since\begin{align*}(n+m_2)-\rank(H)=(n+m_2)-(\rank(H_{short})+\rank(I_{m_2\times m_2}))=\\(n+m_2)-(\rank(H_{short})+  m_2)= n-\rank(H_{short})=\rank(G_{short})=\rank(G),\end{align*}
 we obtain that $G$ is a generator matrix for $H$. 
 
 Let $\vect{v}_1$  be a codeword of the smallest weight in $\ker(H_{short})$, then $$d_{min}(\ker(H))\geq wt(\vect{v}_1, \vect{v}_2)\geq wt(\vect{v}_1)=d_{min}(\ker(H_{short})),$$ 
and the lower bound on the minimum distance  follows. 
\end{IEEEproof}

%%%%%%%%%%%%%%%%%%%%%%%%%%%%%%%%%%%%%%%%%%%%%%%%%
\section{Generator Matrices for QC-GLDPC Codes}\label{sec:genGLDPC}
%%%%%%%%%%%%%%%%%%%%%%%%%%%%%%%%%%%%%%%%%%%%%%%%%%
In this section, we construct generalized LDPC codes directly and also by using pre-lifting~\cite{9762914}, based on  results developed in~\cite{lom18b}  that are likely to have good performance,  and show how we can find polynomial generator matrices for them.  

We start from the general class of $2N\times n_vN$ QC-GLDPC codes based on the all-one protograph,  and give examples for  $n_v=6, 7, 15$. Therefore, we can take, w.l.o.g., the following constraint matrix %$$H\defeq \begin{bmatrix} 1&1&1&\ldots & 1\\ 1& x^{i_2}& x^{i_3}&\ldots&x^{i_{n_v}}\end{bmatrix}.$$
\vspace{-5pt}
\begin{align}\label{matrix2byn}
H(x)\defeq \begin{bmatrix} 1&1&1&\cdots & 1\\ 1& x^{i_2}& x^{i_3}&\cdots&x^{i_{n_v}}\end{bmatrix}.
\end{align}
%Using the sum $1+x^{i_2}+\ldots x^{i_{n_v}}$ and the conditions in paper~\cite{Unified}, we can choose the exponents such that we obtain girth 8 or 12.   
We remark that conditions were given in \cite{9762914} for necessary and sufficient conditions on the exponents $i_2,i_3,\ldots,i_{n_v}$ to achieve a desired girth. For example, the girth is 4 if and only if there are at least 2 exponents that are equal (modulo $N$). 

\subsection{Direct Generalization of One Constraint} We consider first the case where the first constraint ($N$ check nodes) are all simple nodes (single parity-checks) and the second constraint node is a general code with a given parity-check matrix.

\begin{corollary} \label{thm-directly} Let $H(x)$ be a $2\times n_v$ polynomial constraint matrix like in \eqref{matrix2byn}   
and let $h_{GC}=\begin{bmatrix} M_{n\times m}&I_{n\times n}\end{bmatrix} $ be the parity-check matrix for the constraint code in systematic form, where $M$ is an $n\times m$ matrix, $I$ is the $n\times n$ identity matrix, and  $n_v=m+n.$
Let $H_{GC}$ be the parity-check matrix  of the overall GLDPC code based on $H$ and $h_{GC}$, as   
\begin{align}\label{HGC}
H_{GC}&\defeq \left[\begin{array}{cccc|ccc}  1& x^{i_2}&\cdots& x^{i_m}& x^{i_{m+1}}&\cdots&x^{i_{n_v}} \\\hline &&M&&&I&\end{array}\right].\end{align}
Let $ H_{short}\defeq \begin{bmatrix}   f_1&\cdots  &f_m\end{bmatrix}\defeq \begin{bmatrix} 1& x^{i_2}&\cdots& x^{i_m} \end{bmatrix} +\begin{bmatrix}  x^{i_{m+1}}&\cdots&x^{i_{n_v}}\end{bmatrix} M$
and let $G_{short}$, $G'_{short}$ be the  matrices%$(2m-1)N\times mN$ matrix   
{\small  \begin{align}\label{Gshort} G_{short}=\left[\begin{array}{ccccc}  f_m^\tr &0&\cdots &0& f_1^\tr\\ 0&f_m^\tr &\cdots &0& f_2^\tr\\
 \vdots&\vdots&\cdots&\vdots&\vdots\\0 &0 &\cdots &f_m^\tr& f_{m-1}^\tr\\\hline
 f^\tr &0&\cdots & 0& 0\\  0 &f^\tr &\cdots &0 &  0\\ \vdots &\vdots&\cdots&\vdots&\vdots\\ 0 & 0& \cdots & 0&f^\tr\end{array}\right],
 G'_{short}=\left[\begin{array}{ccccc}  \left(\frac{f_m}{g(x)}\right)^\tr &0&\cdots &0& \left(\frac{f_{1}}{g(x)}\right)^\tr\\ 0&\left(\frac{f_m}{g(x)}\right)^\tr &\cdots &0& \left(\frac{f_{2}}{g(x)}\right)^\tr\\ \vdots&\vdots&\cdots&\vdots&\vdots\\0 &0 &\cdots &\left(\frac{f_m}{g(x)}\right)^\tr& \left(\frac{f_{m-1}}{g(x)}\right)^\tr\\\hline
 %f^\tr &0&\cdots & 0& 0\\  0 &f^\tr &\cdots &0 &  0\\ \vdots &\vdots&\cdots&\vdots&\vdots\\ 
 0 & 0& \cdots & 0&f^\tr\end{array}\right], \end{align}} %forms a  generator matrix for $\ker H_{short}$, %= \ker \begin{bmatrix}   f_1&\cdots  &f_m\end{bmatrix}$, 
 where $g(x)$ is defined as  $f(x)g(x)=x^N+1$, with $\gcd(f_1, \ldots, f_m, x^N+1)=g(x)$.\footnote{We need  to have $\gcd(f_m, x^N+1) =g(x)$ in $\Ftwo[x]$ (and not a larger degree polynomial), otherwise, we choose another $f_i$  that satisfies $\gcd(f_i, x^N+1) =g(x)$ to be on the diagonal part of the matrix above. Also, if $g(x)=1$, then $f(x)=0$ and the last $m-1$ rows are omitted. }

Then, $G_{short}$ and $G'_{short}$ are  generator matrices for $\ker(H_{short})$ and  the matrices $$  G= \left[\begin{array}{c|c} G_{short}& G_{short}M^\tr \end{array}\right] ,   G'= \left[\begin{array}{c|c} G'_{short}& G'_{short}M^\tr \end{array}\right] $$ are  generator matrices for $\ker(H_{GC})$. 
 \end{corollary}
\begin{IEEEproof} 
After row operations on \eqref{HGC}, we obtain a row-equivalent  matrix, which is a parity-check matrix for the same code, as
\begin{align*} H_{GC}&\sim \left[\begin{array}{c|c} \begin{bmatrix} 1& x^{i_2}&\cdots& x^{i_m} \end{bmatrix} +\begin{bmatrix}  x^{i_{m+1}}&\cdots&x^{i_{n_v}}\end{bmatrix} M&\vect{0} \\\hline M&I\end{array}\right]\sim  \left[\begin{array}{c|c} \begin{matrix}   f_1&\cdots  &f_m\end{matrix}&\vect{0} \\\hline M&I\end{array}\right] % =\left[\begin{array}{c|c} H_{short}&\vect{0} \\\hline M&I\end{array}\right] 
\end{align*}
We now apply  Theorem~\ref{thm-short} for $H_{short}\defeq \begin{bmatrix}   f_1&\cdots  &f_m\end{bmatrix}$, then Theorem~\ref{case1}, Remark~\ref{case2}, and Lemma~\ref{lemma:codeword:construction:2},   to obtain the generator matrix $G_{short}$ as in the statement.

Finally, we apply   Theorem~\ref{rank} to compute the ranks of $G_{short}$ and $H_{short}$ and show that they are correctly related, where
\begin{align*} \rank (H_{short})&= N-\deg {f(x)} =  \deg g(x),\\
%and the rank of the resulting generating matrix is: 
\rank (G_{short})&=(m-1)(N - \deg {f(x)})+m(N- \deg g(x))\\&= (m-1)(2N - \deg {f(x)} -\deg f(x)) +(N- \deg g(x)) \\&=mN-\deg g(x) =mN-\rank (H_{short}),\textrm{ and}\\\rank (G'_{short})&=(m-1)N +N-\deg g(x))\\&= (m-1)N + \deg {f(x)} \\&=mN-(N-\deg(f(x))=mN-\rank (H_{short}).
\end{align*}
Therefore,  $G_{short}$ and $G'_{short}$  are generator matrices for $H_{short}$. 
Following Theorem~\ref{thm-short}, the matrices $G$ and $G'$   form generator matrices for $\ker(H_{GC})$. 
 %$ G= \left[\begin{array}{c|c} G_{short}& G_{short}M^\tr \end{array}\right].$
%where $ H_{short}\defeq \begin{bmatrix}   f_1&\cdots  &f_m\end{bmatrix}\defeq \begin{bmatrix} 1& x^{i_2}&\cdots& x^{i_m} \end{bmatrix} +\begin{bmatrix}  x^{i_{m+1}}&\cdots&x^{i_{n_v}}\end{bmatrix} M.$
 \end{IEEEproof}

\begin{remark} An alternative way to generalize $H$ is   
\begin{align*}
H'_{GC}&\defeq \left[\begin{array}{cccc|ccc}  1& x^{i_2}&\cdots& x^{i_m}& x^{i_{m+1}}&\cdots&x^{i_{n_v}} \\\hline &&M&&&I&\end{array}\right]\\&\sim\left[\begin{array}{ccc|ccc}  1&  \cdots& 1&1& \cdots&1 \\\hline  &M^{\uparrow\{1,x^{-i_2},\cdots, x^{-i_m}\}}&&&I^{\uparrow\{x^{-i_{m+1}}, \cdots, x^{-i_{n_v}}\}}&\end{array}\right], 
\end{align*}
where $M^{\uparrow\{1,x^{-i_2},\ldots, x^{-i_m}\}}$ and $I^{\uparrow\{x^{-i_{m+1}},\ldots, x^{-i_{n_v}}\}}$ are the generalized (expanded) matrices of $M$ and $I$ respectively, lifted with the circulant order shown in the exponent, according to their sizes.
Note that the two matrices  $H_{GC}$ and $H'_{GC}$  are column-equivalent (equivalent after performing column operations, like dividing by a monomial) which means that they give two equivalent codes, so without any loss of generality, we  work with $H_{GC}$ in the sequel for simplicity. 
\end{remark}

%\begin{remark} The minimum weight  of the rows of the matrix $G$ is  an upper bound to the minimum distance of the GLDPC code, while the minimum distance of the code $\ker (H_{short})$ is a lower bound for the minimum distance of the code. \end{remark} 

%\subsection{Case $n_v=6$} \label{case6}  
\begin{example}\label{example1-GLDPC} Let $n_v=6$, and
%We consider first the case where the first constraint ($N$ check nodes) are all simple nodes and the second constraint node is a shortened Hamming code with parity-check matrix 
%$h_{GC}$ and the parity-check matrix $H_{GC}$ of the GLDPC code below. 
%\vspace{-5pt}
\begin{align}\label{shortHamming}
h_{GC}\defeq \begin{bmatrix} 1&1&0&1&0&0\\1&0&1&0&1&0\\ 0&1&1&0&0&1\end{bmatrix}, ~ %\end{equation}   
%Therefore,\vspace{-5pt}
%\begin{align} \label{gc_4x6_from_nv6}
%H_{GC}\defeq \left[\begin{array}{cccccc}  1&1&1&1&1& 1\\\hline 1&x^{-i_2}&0&x^{-i_4}&0&0\\1&0&x^{-i_3}&0&x^{-i_5}&0\\ 0&x^{-i_2}&x^{-i_3}&0&0&x^{-i_6}\end{array}\right]. 
%\end{align}
%Since  the GLDPC code with parity-check matrix $H_{GC}$ as in \eqref{shortHamming} %{gc_4x6_from_nv6}
%%\vspace{-5pt}
%%\begin{align*}
%%H_{GC}\defeq \left[\begin{array}{cccccc}  1&1&1&1&1& 1\\\hline1&x^{i_2}&0&x^{i_4}&0&0\\1&0&x^{i_3}&0&x^{i_5}&0\\ 0&x^{i_2}&x^{i_3}&0&0&x^{i_6}\end{array}\right]
%%\end{align*} 
%is equivalent to the one with parity-check matrix 
%\vspace{-5pt}
%\begin{align*}
H_{GC}\defeq \left[\begin{array}{ccc|ccc}  1& x^{i_2}& x^{i_3}&x^{i_4}&x^{i_5}&x^{i_6}\\\hline1&1&0&1&0&0\\1&0&1&0&1&0\\ 0&1&1&0&0&1\end{array}\right] 
%\end{align}
%we  work with this matrix instead. 
%
%
%
%\begin{align*}H'_{GC} %\defeq \left[\begin{array}{cccccc}  1&1&1&1&1& 1\\\hline 1&x^{i_2}&0&x^{i_4}&0&0\\1&0&x^{i_3}&0&x^{i_5}&0\\ 0&x^{i_2}&x^{i_3}&0&0&x^{i_6}\end{array}\right] \sim\begin{bmatrix}  1&1&1&1&1& 1\\x^{-i_4}&x^{i_2-i_4}&0&1&0&0\\x^{-i_5}&0&x^{i_3-i_5}&0&1&0\\ 0&x^{i_2-i_6}&x^{i_3-i_6}&0&0&1\end{bmatrix} \sim $$ 
%&\sim \left[\begin{array}{c|c}  \begin{bmatrix}1&x^{i_2}&x^{i_3}\end{bmatrix}+\begin{bmatrix} x^{i_4}&x^{i_5}&x^{i_6}\end{bmatrix}M &0\\ M&I
%\end{array}\right]\\&
%\sim \left[\begin{array}{ccc|ccc}   1+x^{i_4}+x^{i_5}&x^{i_2}+x^{i_4}+x^{i_6}&x^{i_3}+x^{i_5}+x^{i_6}&0&0& 0\\\hline 1&1&0&1&0&0\\1&0&1&0&1&0\\ 0&1&1&0&0&1\end{array}\right] \\&
\sim \left[\begin{array}{c|c} \begin{matrix}   f_1&f_2 &f_3\end{matrix}&\vect{0} \\\hline M&I\end{array}\right],  \end{align}
where $f_1\defeq 1+x^{i_4}+x^{i_5}$, $f_2\defeq x^{i_2}+x^{i_4}+x^{i_6}$, and $f_3\defeq x^{i_3}+x^{i_5}+x^{i_6}$. Let $g(x)\defeq \gcd(f_1,f_2,f_3, x^N+1)$ and $f(x)$ be such that  $f(x)g(x)=x^N+1$. Following Theorem~\ref{thm-directly},
 the matrix   
 $$G= \left[\begin{array}{c|c} G_{short}& G_{short}M^\tr \end{array}\right]= \left[\begin{array}{ccc|ccc} f_3^\tr &0& f_1^\tr&f_3^\tr & f_1^\tr +f_3^\tr&f_1^\tr\\ 
 0&f_3^\tr& f_2^\tr&f_3^\tr&f_2^\tr&f_2^\tr +f_3^\tr \\ f^\tr &0&0&f^\tr&f^\tr&0\\ 0& f^\tr& 0& f^\tr & 0 &f^\tr\\0 &0&f^\tr&0&f^\tr&f^\tr  \end{array}\right]$$
 forms a generator matrix for $\ker(H_{GC})$. 

  For example, for $N=79$, and  exponents  $[i_2, i_3, i_4, i_5, i_6] =[54, 66, 71, 55, 69]$ given in~\cite{lom18b}, for  which $H$ has girth 12,\footnote{How to choose exponents to achieve the desired girth in the constraint matrix is given in \cite{9762914}.}
we compute $f_i$ and obtain 
$f_1=1+x^{71}+x^{55}, f_2=x^{54}+x^{71}+ x^{69}, f_3=x^{66}+x^{55}+x^{69}.
$
Since $\gcd(f_3, x^{79}+1)=1$ in $\Ftwo[x]$, the last three rows of the matrix $G$ are omitted. %The  matrix  $$G_{syst}=\begin{bmatrix} 1 &0& f_1f_3^{-1}&1 & f_1f_3^{-1} +1&f_1f_3^{-1}\\ 
% 0&1& f_2f_3^{-1}&1&f_2f_3^{-1}&f_2f_3^{-1} +1 \end{bmatrix}$$
% is a systematic generator matrix for the code, but it is very dense, while 
The polynomial generator matrix $G$ has each row of  weight 16, which is  the minimum distance of the code. Therefore, we obtain a $[474, 158, 16]$ code with a generator matrix of the minimum possible weight.  
 \end{example} 
\begin{example} \label{example2-GLDPC} Let $n_v=7$ and 
% We consider that the component codes will be the Hamming code with parity check matrix 
%\vspace{-6pt}
%\begin{equation}\label{Hamming}
\begin{align*} h_{GC}\defeq \begin{bmatrix} 1&1&1&0&1&0&0\\1&1&0&1&0&1&0\\ 1&0&1&1&0&0&1\end{bmatrix}=\begin{bmatrix} M&I\end{bmatrix},   %\end{equation}
%(or any other parity-check matrix of a Hamming $(7,4,3)$ code).
%By generalizing one constraint,  we get the parity-check matrix 
%\vspace{-10pt}
H_{GC} %\defeq& %\left[\begin{array}{cccc|ccc}  1&x^{i_2}&x^{i_3}&x^{i_4}&x^{i_5}&x^{i_6}&x^{i_7}\\\hline1&1&1&0&1&0&0\\1&1&0&1&0&1&0\\ 1&0&1&1&0&0&1
%\end{array}\right]
%\sim \left[\begin{array}{c|c}  \begin{bmatrix}1&x^{i_2}&x^{i_3}&x^{i_4}\end{bmatrix}+\begin{bmatrix} x^{i_5}&x^{i_6}&x^{i_7}\end{bmatrix}M &0\\ M&I
%\end{array}\right] \\
%&\left[\begin{array}{cccc|ccc} 1+x^{i_5}+x^{i_6}+x^{i_7}  &x^{i_2}+x^{i_5}+x^{i_6}&x^{i_3}+x^{i_5}+x^{i_7}&x^{i_4}+x^{i_6}+x^{i_7}&0&0& 0\\\hline 1 &1&1&0&1&0&0\\1&1&0&1&0&1&0\\ 1&0&1&1&0&0&1\end{array}\right]
\sim  \left[\begin{array}{c|c} \begin{matrix}   f_1&f_2 &f_3&f_4\end{matrix}&\vect{0} \\\hline M&I\end{array}\right], \end{align*} where  $ f_1=    1+x^{i_5}+x^{i_6}+x^{i_7}, f_2= x^{i_2}+x^{i_5}+x^{i_6}, f_3=x^{i_3}+x^{i_5}+x^{i_7}, f_4=x^{i_4}+x^{i_6}+x^{i_7}.$
Following Theorem~\ref{thm-directly}, 
 the matrix  $G_{GC}$ below forms the main part of a polynomial generator matrix for the generalized code $\ker(H_{GC})$,   
%code $\ker(H_{short}=\ker(\begin{bmatrix}   f_1&f_2 &f_3&f_4\end{bmatrix})$, 
to which rows containing the polynomial $f(x)$ must added if the rank is not full, 
 $$G_{GC}= \left[\begin{array}{cccc|ccc} 
 f_4^\tr  &  0       &0       & f_1^\tr   &f_4^\tr    & f_1^\tr +f_4^\tr   &f_1^\tr+f_4^\tr \\ 
 0          & f_4^\tr  &0       & f_2^\tr   &f_4^\tr    &f_2^\tr    +f_4^\tr             &f_2^\tr  \\ 
 0          &0         &f_4^\tr & f_3^\tr   &f_4^\tr    &f_3^\tr                 &f_3^\tr +f_4^\tr \end{array}\right].$$%=\begin{bmatrix} G&GM^\tr\end{bmatrix}. $$ 
% giving $$ \left[\begin{array}{c|c} \begin{matrix}   f_1&\cdots  &f_m\end{matrix}&\vect{0} \\\hline M&I\end{array}\right] = \begin{bmatrix}
% f_4^\tr  &  0       &0       & f_1^\tr    \\ 
% 0          & f_4^\tr  &0       & f_2^\tr     \\ 
% 0          &0         &f_4^\tr & f_3^\tr   \end{bmatrix},$$   a generator matrix for the . %We can create a similar matrix with any $f_i$ on the diagonal of a $3\times 3$ submatrix. 
%If the gcd between any of the $f_i$ and $x^N+1$ is 1, then $G_{b}$ is a generator matrix, otherwise, the rank of $G_{b}$ is lower than the dimension of the GLDPC code given by $H'_{GC}$, so  we would need to add a few naturally occurring codewords to obtain a generator matrix. We do this  using the method in~\cite{sgm22}. For example, suppose that $\gcd(f_4,x^N+1)=x+1$ and let $f(x)$ defined as $(x+1)f(x)=x^N+1$. Then, the codewords of even weight in the code given by the  parity-check matrix $h_{GC}$  multiplied by the  $f(x)$ will be codewords in the GLDPC given by $H$, i.e., 
%$\left(f, f, 0,0,0, f,f\right), \left(0, f,0, f,f,f, 0\right), \left(f, 0, 0,  f,f,0, f\right). $
%Any two of them are linearly independent and therefore, can be added to $G_{b}$ to increase the rank. Since these codewords have large weight, we can form linear combinations between these codewords and the rows of $G_{b}$ to create codewords that add to the rank but are less dense. 
The weight (16 in this case) of the rows of the matrix $G_{GC}$ is  an upper bound to the minimum distance of the GLDPC code.

 For example, let $N=68$ and set the exponents $[i_2, i_3, i_4, i_5, i_6, i_7] =[61, 49, 44, 1, 46, 14]$, such that $H$ has girth 12~\cite{lom18b}.  %How to choose these exponents to achieve the desired girth is given in our paper. 
We compute $f_i$ and obtain 
$f_1=1+x+ x^{46}+x^{14}, f_2=x^{61}+x+ x^{46}, f_3=x^{49}+x+x^{14}, f_4=x^{44}+x^{46}+x^{14}.
$
 Since $\gcd(f_4, x^{68}+1)=1$ in $\Ftwo[x]$, $f_4$ is invertible and the matrix $G$ is a polynomial generator matrix of the code.  
Therefore, we obtain a $[476, 204, 16]$ code with a generator matrix of the minimum possible weight. We denote this code $C_1$ and will present simulation results for it later in Section \ref{sec:sim}.
% We observe that the codewords can be viewed as $\vect{v}^\tr=(\vect{v}^\tr_1, \vect{v}^\tr_2)$ with $$\vect{v}_1\in \ker(\begin{matrix}   f_1&f_2 &f_3&f_4\end{matrix}), \vect{v}_2=M\vect{v}_1 ,$$
% so  $d_{min}(\ker(\begin{matrix}   f_1&f_2 &f_3&f_4\end{matrix}) )$ is a lower bound for the minimum distance of the code, and a generator matrix $G$ for $\ker(\begin{matrix}   f_1&f_2 &f_3&f_4\end{matrix}) $ is used to create the 
% generator matrix for the code as $\begin{bmatrix} G & GM^\tr\end{bmatrix}$. 
 \end{example}
%\subsection{Case $n_v=15$}\label{case15} 
This is demonstrated for a code with constraint degree $n_v = 15$ in Appendix \ref{app:ham15}.

%
%
% and minimum distance 6 (only an increase of 2 over the original), with a decrease of rank ($404$ decreased to $310$). The reason for the improvement not being significant is that  $N$ is too small.   
\subsection{Direct Generalization of Both  Constraints} The following theorem considers the case when both  constraints are generalized.  
\begin{corollary} \label{thm-directly-both} Let $H$ be a $2\times n_v$ polynomial constraint matrix like in \eqref{matrix2byn}   
and let $h_{GC}=\begin{bmatrix} M_{n\times m}&I_{n\times n}\end{bmatrix} $ and $h'_{GC}=\begin{bmatrix} M_1&M_2 \end{bmatrix}$, where $M$, $M_1$, and $M_2$ are $n\times m$,  $k\times m$, and $k \times n$ matrices, respectively, $I$ is the $n\times n$ identity matrix, and  $n_v=m+n. $ Here, $h'_{GC}$ and $h_{GC}$ correspond to the two generalized constraints. Let $H_{GC}$ be the overall parity-check matrix  of the GLDPC code based on $H$, $h'_{GC}$, and $h_{GC}$ as
\begin{align}\label{HGC-2}
H_{GC}&\defeq \left[\begin{array}{c|c}  M_1^{\uparrow \{\begin{matrix}1& x^{i_2}&\cdots& x^{i_m}\end{matrix}\}}&M_2^{\uparrow \{\begin{matrix} x^{i_{m+1}}&\cdots&x^{i_{n_v}}\end{matrix}\}} \\\hline M&I\end{array}\right].\end{align}
Let $ H_{short}\defeq  M_1^{\uparrow \{\begin{matrix}1& x^{i_2}&\cdots& x^{i_m}\end{matrix}\}}  +M_2^{\uparrow \{\begin{matrix} x^{i_{m+1}}&\cdots&x^{i_{n_v}}\end{matrix}\}}M$, a $k\times m$ matrix), and let $G_{short}$ be  a generator matrix for  $\ker(H_{short})$.
% \begin{align}\label{Gshort} G_{short}=\left[\begin{array}{ccccc}  f_m^\tr &0&\cdots &0& f_1^\tr\\ 0&f_m^\tr &\cdots &0& f_2^\tr\\
% \vdots&\vdots&\cdots&\vdots&\vdots\\0 &0 &\cdots &f_m^\tr& f_{m-1}^\tr\\\hline
% g^\tr &0&\cdots & 0& 0\\  0 &g^\tr &\cdots &0 &  0\\ \vdots &\vdots&\cdots&\vdots&\vdots\\ 0 & 0& \cdots & 0&g^\tr\end{array}\right],\end{align} %forms a  generator matrix for $\ker H_{short}$, %= \ker \begin{bmatrix}   f_1&\cdots  &f_m\end{bmatrix}$, 
% where $g(x)$ is defined as  $f(x)g(x)=x^N+1$, with $\gcd(f_1, \ldots, f_m, x^N+1)=f(x)$.\footnote{We need  to have $\gcd(f_m, x^N+1) =f(x)$ in $\Ftwo[x]$ (and not a larger degree polynomial), otherwise, we choose another $f_i$  that satisfies $\gcd(f_i, x^N+1) =f(x)$ to be on the diagonal part of the matrix above.} 
%\footnote{If $f(x)=1$, then $g(x)=0$ and the last $m-1$ rows are omitted. }
Then the matrix $  G= \left[\begin{array}{c|c} G_{short}& G_{short}M^\tr \end{array}\right] $ is  a generator matrix for $\ker(H_{GC})$ (to which some extra vectors need to be included, if it is not full rank).  
 \end{corollary}
\begin{IEEEproof} 
After performing row operations on \eqref{HGC-2} that use the identity $I$ to make a zero matrix above it, we obtain the following row-equivalent  matrix which is a parity-check matrix for the same code 
\begin{align*} H_{GC}&\sim \left[\begin{array}{c|c} H_{short}&\vect{0} \\\hline M&I\end{array}\right]. \end{align*}
%We apply   Theorem~\ref{case1},  to obtain the first $m-1$ rows of $G_{short}$ and check directly that the lower part of the matrix is formed by codewords. 
%We apply   Theorem~\ref{thm-ranks} to compute the ranks of $G_{short}$ and $H_{short}$ and show that they are correctly related. 
%\begin{align*} \rank (H_{short})&= N-\deg {f(x)} =  \deg g(x)\\
%%and the rank of the resulting generating matrix is: 
%\rank (G_{short})&=(m-1)(N - \deg {f(x)})+m(N- \deg g(x))\\&= (m-1)(2N - \deg {f(x)} -\deg f(x)) +(N- \deg g(x)) \\&=mN-\deg g(x) =mN-\rank (H_{short})\end{align*}
%
%Therefore,  $G_{short}$ is  a generator matrix for $H_{short}$. 
Following Theorem~\ref{thm-short}, the matrix $ G$   forms  a generator matrix for $\ker(H_{GC})$. 
 %$ G= \left[\begin{array}{c|c} G_{short}& G_{short}M^\tr \end{array}\right].$
%where $ H_{short}\defeq \begin{bmatrix}   f_1&\cdots  &f_m\end{bmatrix}\defeq \begin{bmatrix} 1& x^{i_2}&\cdots& x^{i_m} \end{bmatrix} +\begin{bmatrix}  x^{i_{m+1}}&\cdots&x^{i_{n_v}}\end{bmatrix} M.$
 \end{IEEEproof}
%\begin{example}\label{example1-GLDPC-2} 
{\bf Example \ref{example1-GLDPC} (cont.).} Let $n_v=6$, and 
%We consider first the case where the first constraint ($N$ check nodes) are all simple nodes and the second constraint node is a shortened Hamming code with parity-check matrix 
%$h_{GC}$ and the parity-check matrix $H_{GC}$ of the GLDPC code below. 
%\vspace{-5pt}
{\begin{align*}&h_{GC}\defeq \begin{bmatrix} M&I\end{bmatrix}=\begin{bmatrix} 1&1&0&1&0&0\\1&0&1&0&1&0\\ 0&1&1&0&0&1\end{bmatrix}, ~h'_{GC}=\begin{bmatrix} M_1&M_2 \end{bmatrix},\textrm{ and}%\end{equation}   
%Therefore,\vspace{-5pt}
%\begin{align} \label{gc_4x6_from_nv6}
%H_{GC}\defeq \left[\begin{array}{cccccc}  1&1&1&1&1& 1\\\hline 1&x^{-i_2}&0&x^{-i_4}&0&0\\1&0&x^{-i_3}&0&x^{-i_5}&0\\ 0&x^{-i_2}&x^{-i_3}&0&0&x^{-i_6}\end{array}\right]. 
%\end{align}
%Since  the GLDPC code with parity-check matrix $H_{GC}$ as in \eqref{shortHamming} %{gc_4x6_from_nv6}
%%\vspace{-5pt}
%%\begin{align*}
%%H_{GC}\defeq \left[\begin{array}{cccccc}  1&1&1&1&1& 1\\\hline1&x^{i_2}&0&x^{i_4}&0&0\\1&0&x^{i_3}&0&x^{i_5}&0\\ 0&x^{i_2}&x^{i_3}&0&0&x^{i_6}\end{array}\right]
%%\end{align*} 
%is equivalent to the one with parity-check matrix 
%\vspace{-5pt} \left[\begin{array}{c|c} & \\\hline M&I\end{array}\right]%\begin{align*}
\\ &H_{GC}\defeq \left[\begin{array}{c|c}  M_1^{\uparrow \{\begin{matrix}1& x^{i_2}& x^{i_3} \end{matrix}\}}&M_2^{\uparrow \{\begin{matrix} x^{i_4}&x^{i_5}&x^{i_6}\end{matrix}\}}\\\hline M&I\end{array}\right] \sim \left[\begin{array}{c|c} H_{short}&\vect{0} \\\hline M&I\end{array}\right], \end{align*}}
%we  work with this matrix instead. 
%
%
%
%\begin{align*}H'_{GC} %\defeq \left[\begin{array}{cccccc}  1&1&1&1&1& 1\\\hline 1&x^{i_2}&0&x^{i_4}&0&0\\1&0&x^{i_3}&0&x^{i_5}&0\\ 0&x^{i_2}&x^{i_3}&0&0&x^{i_6}\end{array}\right] \sim\begin{bmatrix}  1&1&1&1&1& 1\\x^{-i_4}&x^{i_2-i_4}&0&1&0&0\\x^{-i_5}&0&x^{i_3-i_5}&0&1&0\\ 0&x^{i_2-i_6}&x^{i_3-i_6}&0&0&1\end{bmatrix} \sim $$ 
%&\sim \left[\begin{array}{c|c}  \begin{bmatrix}1&x^{i_2}&x^{i_3}\end{bmatrix}+\begin{bmatrix} x^{i_4}&x^{i_5}&x^{i_6}\end{bmatrix}M &0\\ M&I
%\end{array}\right]\\&
%\sim \left[\begin{array}{ccc|ccc}   1+x^{i_4}+x^{i_5}&x^{i_2}+x^{i_4}+x^{i_6}&x^{i_3}+x^{i_5}+x^{i_6}&0&0& 0\\\hline 1&1&0&1&0&0\\1&0&1&0&1&0\\ 0&1&1&0&0&1\end{array}\right] \\&
%\sim \left[\begin{array}{c|c} \begin{matrix}   f_1&f_2 &f_3\end{matrix}&\vect{0} \\\hline M&I\end{array}\right],  
where $h'_{GC}$ is a column permutation of $h_{GC}$,\footnote{Note that taking one constraint code as a column permutation of another is a well established technique to obtain good GLDPC codes, see, e.g., \cite{lrc08,molc21}.} and $$ H_{short}\defeq  M_1^{\uparrow \{\begin{matrix}1& x^{i_2}& x^{i_3}\end{matrix}\}}  + M_2^{\uparrow \{\begin{matrix} x^{i_4}&x^{i_5}&x^{i_6}\end{matrix}\}}M.$$ %$f_1\defeq 1+x^{i_4}+x^{i_5}$, $f_2\defeq x^{i_2}+x^{i_4}+x^{i_6}$, and $f_3\defeq x^{i_3}+x^{i_5}+x^{i_6}$. 
For example, for $h'_{GC}=\begin{bmatrix}I& M\end{bmatrix}$,
{\small \begin{align*} H_{GC} &= \left[\begin{array}{ccc|ccc} 1& 0&0& x^{i_4}& x^{i_5}&0\\0& x^{i_2}& 0& x^{i_4}&0 &x^{i_6}\\0&0&x^{i_3}&0&x^{i_5}&x^{i_6} \\\hline 1&1&0&1&0&0\\1&0&1&0&1&0\\ 0&1&1&0&0&1\end{array}\right]\sim  \left[\begin{array}{ccc|ccc}1+ x^{i_4}+ x^{i_5} & x^{i_4}&x^{i_5}& 0& 0&0\\x^{i_4}& x^{i_2}+x^{i_4}+x^{i_6}& x^{i_6}&0&0 &0\\x^{i_5}&x^{i_6}& x^{i_3}+x^{i_5}+x^{i_6}&0&0&0 \\\hline 1&1&0&1&0&0\\1&0&1&0&1&0\\ 0&1&1&0&0&1\end{array}\right],\\
H_{short}&\defeq  I^{\uparrow \{\begin{matrix}1& x^{i_2}& x^{i_3}\end{matrix}\}}  + M^{\uparrow \{\begin{matrix} x^{i_4}&x^{i_5}&x^{i_6}\end{matrix}\}}M =  \left[\begin{array}{ccc}1+ x^{i_4}+ x^{i_5} & x^{i_4}&x^{i_5}\\x^{i_4}& x^{i_2}+x^{i_4}+x^{i_6}& x^{i_6}\\x^{i_5}&x^{i_6}& x^{i_3}+x^{i_5}+x^{i_6}\end{array}\right].\end{align*}}
 The matrix $H_{short}$ is not invertible, therefore, $\ker(H_{short})\neq \emptyset$, and we use Lemma~\ref{lemma:codeword:construction:2} to obtain the generator matrix of very low rank.  We could instead  disregard one of the first three rows, and form a better code.\hfill $\Box$

\noindent{\bf Example \ref{example2-GLDPC} (cont.).} In this example, we demonstrate the effect of the choices of $M_1$ and $M_2$. Let $n_v=7$ and 
% We consider that the component codes will be the Hamming code with parity check matrix 
%\vspace{-6pt}
%\begin{equation}\label{Hamming}
\begin{align*} &h_{GC}\defeq \begin{bmatrix} 1&1&1&0&1&0&0\\1&1&0&1&0&1&0\\ 1&0&1&1&0&0&1\end{bmatrix}=\begin{bmatrix} M&I\end{bmatrix}, \quad  h'_{GC}=\begin{bmatrix} M_1&M_2\end{bmatrix}, \textrm{ and}%\end{equation}   
%Therefore,\vspace{-5pt}
%\begin{align} \label{gc_4x6_from_nv6}
%H_{GC}\defeq \left[\begin{array}{cccccc}  1&1&1&1&1& 1\\\hline 1&x^{-i_2}&0&x^{-i_4}&0&0\\1&0&x^{-i_3}&0&x^{-i_5}&0\\ 0&x^{-i_2}&x^{-i_3}&0&0&x^{-i_6}\end{array}\right]. 
%\end{align}
%Since  the GLDPC code with parity-check matrix $H_{GC}$ as in \eqref{shortHamming} %{gc_4x6_from_nv6}
%%\vspace{-5pt}
%%\begin{align*}
%%H_{GC}\defeq \left[\begin{array}{cccccc}  1&1&1&1&1& 1\\\hline1&x^{i_2}&0&x^{i_4}&0&0\\1&0&x^{i_3}&0&x^{i_5}&0\\ 0&x^{i_2}&x^{i_3}&0&0&x^{i_6}\end{array}\right]
%%\end{align*} 
%is equivalent to the one with parity-check matrix 
%\vspace{-5pt} \left[\begin{array}{c|c} & \\\hline M&I\end{array}\right]%\begin{align*}
\\ &H_{GC}\defeq \left[\begin{array}{c|c}  M_1^{\uparrow \{\begin{matrix}x^{i_1}& x^{i_2}& x^{i_3} & x^{i_4}\end{matrix}\}}&M_2^{\uparrow \{\begin{matrix} x^{i_5}&x^{i_6}&x^{i_7}\end{matrix}\}}\\\hline M&I\end{array}\right] \sim \left[\begin{array}{c|c} H_{short}&\vect{0} \\\hline M&I\end{array}\right], %\end{align}
%we  work with this matrix instead. 
%
%
%
%\begin{align*}H'_{GC} %\defeq \left[\begin{array}{cccccc}  1&1&1&1&1& 1\\\hline 1&x^{i_2}&0&x^{i_4}&0&0\\1&0&x^{i_3}&0&x^{i_5}&0\\ 0&x^{i_2}&x^{i_3}&0&0&x^{i_6}\end{array}\right] \sim\begin{bmatrix}  1&1&1&1&1& 1\\x^{-i_4}&x^{i_2-i_4}&0&1&0&0\\x^{-i_5}&0&x^{i_3-i_5}&0&1&0\\ 0&x^{i_2-i_6}&x^{i_3-i_6}&0&0&1\end{bmatrix} \sim $$ 
%&\sim \left[\begin{array}{c|c}  \begin{bmatrix}1&x^{i_2}&x^{i_3}\end{bmatrix}+\begin{bmatrix} x^{i_4}&x^{i_5}&x^{i_6}\end{bmatrix}M &0\\ M&I
%\end{array}\right]\\&
%\sim \left[\begin{array}{ccc|ccc}   1+x^{i_4}+x^{i_5}&x^{i_2}+x^{i_4}+x^{i_6}&x^{i_3}+x^{i_5}+x^{i_6}&0&0& 0\\\hline 1&1&0&1&0&0\\1&0&1&0&1&0\\ 0&1&1&0&0&1\end{array}\right] \\&
%\sim \left[\begin{array}{c|c} \begin{matrix}   f_1&f_2 &f_3\end{matrix}&\vect{0} \\\hline M&I\end{array}\right],  
\end{align*}
where $ H_{short}\defeq  M_1^{\uparrow \{\begin{matrix}x^{i_1}& x^{i_2}& x^{i_3}& x^{i_4}\end{matrix}\}}  + M_2^{\uparrow \{\begin{matrix} x^{i_5}&x^{i_6}&x^{i_7}\end{matrix}\}}M .$ %$f_1\defeq 1+x^{i_4}+x^{i_5}$, $f_2\defeq x^{i_2}+x^{i_4}+x^{i_6}$, and $f_3\defeq x^{i_3}+x^{i_5}+x^{i_6}$. 

%H_{GC} &= \left[\begin{array}{ccc|ccc} 1& 0&0& x^{i_4}& x^{i_5}&0\\0& x^{i_2}& 0& x^{i_4}&0 &x^{i_6}\\0&0&x^{i_3}&0&x^{i_5}&x^{i_6} \\\hline 1&1&0&1&0&0\\1&0&1&0&1&0\\ 0&1&1&0&0&1\end{array}\right]\sim  \left[\begin{array}{ccc|ccc}1+ x^{i_4}+ x^{i_5} & x^{i_4}&x^{i_5}& 0& 0&0\\x^{i_4}& x^{i_2}+x^{i_4}+x^{i_6}& x^{i_6}&0&0 &0\\x^{i_5}&x^{i_6}& x^{i_3}+x^{i_5}+x^{i_6}&0&0&0 \\\hline 1&1&0&1&0&0\\1&0&1&0&1&0\\ 0&1&1&0&0&1\end{array}\right]\\

 The matrices $M_1$ and $M_2$ can be any taken in many different ways. First, consider the choices below and the corresponding $H_{short}$
 {\small \begin{align*} M_1\defeq &\begin{bmatrix} 1&0&0&1\\0&1&0&1\\ 0&0&1&1\end{bmatrix}, M_2\defeq \begin{bmatrix} 1&1&0\\1&0&1\\ 0&1&1\end{bmatrix}, \\
  H_{short}=&  \left[\begin{array}{cccc}x^{i_1}+ x^{i_5}+ x^{i_6} &x^{i_5}+ x^{i_6}& x^{i_5} &x^{i_4}+ x^{i_6}\\x^{i_5}+x^{i_7}& x^{i_2}+ x^{i_5}&x^{i_5}+x^{i_7}&x^{i_4}+x^{i_7}\\  x^{i_6}+x^{i_7}&x^{i_6}&x^{i_3}+x^{i_7}&x^{i_4}+  x^{i_6}+x^{i_7}\end{array}\right]\sim \\& \left[\begin{array}{cccc}x^{i_1}& x^{i_2}&x^{i_3}& x^{i_4}\\x^{i_5}+x^{i_7}& x^{i_2}+ x^{i_5}&x^{i_5}+x^{i_7}&x^{i_4}+x^{i_7}\\x^{i_1}+ x^{i_5}+ x^{i_6} &x^{i_5}+ x^{i_6}& x^{i_5} &x^{i_4}+ x^{i_6} \end{array}\right].\end{align*}}
  %\end{equation}
%(or any other parity-check matrix of a Hamming $(7,4,3)$ code).
%By generalizing one constraint,  we get the parity-check matrix 
%\vspace{-10pt}
%H_{GC} %\defeq& %\left[\begin{array}{cccc|ccc}  1&x^{i_2}&x^{i_3}&x^{i_4}&x^{i_5}&x^{i_6}&x^{i_7}\\\hline1&1&1&0&1&0&0\\1&1&0&1&0&1&0\\ 1&0&1&1&0&0&1
%\end{array}\right]
%\sim \left[\begin{array}{c|c}  \begin{bmatrix}1&x^{i_2}&x^{i_3}&x^{i_4}\end{bmatrix}+\begin{bmatrix} x^{i_5}&x^{i_6}&x^{i_7}\end{bmatrix}M &0\\ M&I
%\end{array}\right] \\
%&\left[\begin{array}{cccc|ccc} 1+x^{i_5}+x^{i_6}+x^{i_7}  &x^{i_2}+x^{i_5}+x^{i_6}&x^{i_3}+x^{i_5}+x^{i_7}&x^{i_4}+x^{i_6}+x^{i_7}&0&0& 0\\\hline 1 &1&1&0&1&0&0\\1&1&0&1&0&1&0\\ 1&0&1&1&0&0&1\end{array}\right]
%\sim  \left[\begin{array}{c|c} \begin{matrix}   f_1&f_2 &f_3&f_4\end{matrix}&\vect{0} \\\hline M&I\end{array}\right], \end{align*} where  $ f_1=    1+x^{i_5}+x^{i_6}+x^{i_7}, f_2= x^{i_2}+x^{i_5}+x^{i_6}, f_3=x^{i_3}+x^{i_5}+x^{i_7}, f_4=x^{i_4}+x^{i_6}+x^{i_7}.$
Following Theorem~\ref{thm-directly}, 
 the matrix  $G_{GC}=\begin{bmatrix} G_{short}&G_{short}M^\tr\end{bmatrix}$  forms the  polynomial generator matrix for the generalized code $\ker(H_{GC})$,  
%code $\ker(H_{short}=\ker(\begin{bmatrix}   f_1&f_2 &f_3&f_4\end{bmatrix})$, 
 where 
 $$G_{short}= \left[\begin{array}{cccc} 
 \Delta_{234}^\tr  &   \Delta_{134}^\tr       & \Delta_{124}^\tr   & \Delta_{123}^\tr    \end{array}\right],$$
  to which extra rows need to be added if the matrix is not full rank.  
  %=\begin{bmatrix} G&GM^\tr\end{bmatrix}. $$ 
% giving $$ \left[\begin{array}{c|c} \begin{matrix}   f_1&\cdots  &f_m\end{matrix}&\vect{0} \\\hline M&I\end{array}\right] = \begin{bmatrix}
% f_4^\tr  &  0       &0       & f_1^\tr    \\ 
% 0          & f_4^\tr  &0       & f_2^\tr     \\ 
% 0          &0         &f_4^\tr & f_3^\tr   \end{bmatrix},$$   a generator matrix for the . %We can create a similar matrix with any $f_i$ on the diagonal of a $3\times 3$ submatrix. 
%If the gcd between any of the $f_i$ and $x^N+1$ is 1, then $G_{b}$ is a generator matrix, otherwise, the rank of $G_{b}$ is lower than the dimension of the GLDPC code given by $H'_{GC}$, so  we would need to add a few naturally occurring codewords to obtain a generator matrix. We do this  using the method in~\cite{sgm22}. For example, suppose that $\gcd(f_4,x^N+1)=x+1$ and let $f(x)$ defined as $(x+1)f(x)=x^N+1$. Then, the codewords of even weight in the code given by the  parity-check matrix $h_{GC}$  multiplied by the  $f(x)$ will be codewords in the GLDPC given by $H$, i.e., 
%$\left(f, f, 0,0,0, f,f\right), \left(0, f,0, f,f,f, 0\right), \left(f, 0, 0,  f,f,0, f\right). $
%Any two of them are linearly independent and therefore, can be added to $G_{b}$ to increase the rank. Since these codewords have large weight, we can form linear combinations between these codewords and the rows of $G_{b}$ to create codewords that add to the rank but are less dense. 

 For example, let  $N=68$ with exponents $ [i_1, i_2, \ldots, i_7] =[ 0, 61, 49, 44, 1, 46, 14 ]$ chosen to give a $2\times 7$ matrix with the associated  Tanner graph having girth 12. We compute $\Delta_{ijk}$ and obtain a weight 52 codeword in $\ker(H_{short})$, which is the possible minimum distance of this code ($34\leq d\leq 52$ from Magma computations). Explicitly,
{\small \begin{align*}\Delta_{234}=& x^{64} + x^{59} + x^{53} + x^{51} + x^{41} + x^{40} + x^{38} + x^{36} + x^{28} + x^{23} + x^{20} + 
    x^{18} + x^{8} + x^{3}, \\
\Delta_{134}=&  x^{64} + x^{63} + x^{60} + x^{47} + x^{45} + x^{39} + x^{36} + x^{28} + 
    x^{25} + x^{23} + x^{15} + x^{3}, \\
\Delta_{124}=&x^{60} + x^{59} + x^{51} + x^{47} + x^{45} + x^{40} + x^{39} + 
    x^{38} + x^{37} + x^{36} + x^{22} + x^{15} + x^{8} + x^{7}, \\ \Delta_{123}=& x^{64} + x^{60} + x^{53} + x^{50} + 
    x^{47} + x^{43} + x^{42} + x^{41} + x^{40} + x^{20} + x^{15} + x^{7} . \end{align*}}Since $\gcd(\Delta_{234}, 
\Delta_{134}, \Delta_{124},  \Delta_{123}, x^{68}+1)=x^4+ 1 $, in $\Ftwo[x]$,  we obtain that $G_{short}$ only has rank $N-4=64$.  We let $f(x)$ be defined such that $f(x)(x^4+1)=x^{68}+1$  in $\Ftwo[x]$, and add to this matrix $ G_{short} $, any two of  the codewords 
$$ (0,g_1,g_2,g_3),(g_3,g_4,g_5, 0), (g_1, 0, g_6,g_4),   (g_2,  g_6,0, g_5),$$
where 
\begin{align*} 
g_1=&f(x) \Delta_{1,3|3,4}   = f(x)( x^{i_3}(x^{i_4}+ x^{i_6}) +x^{i_4+i_5}),\\
g_2=&f(x) \Delta_{1,3|2,4}   = f(x)( x^{i_2}(x^{i_4}+ x^{i_6}) +x^{i_4}( x^{i_5}+x^{i_6}),\\ 
g_3=&f(x) \Delta_{1,3|2,3}   = f(x)( x^{i_2+i_5} +x^{i_3}( x^{i_5}+x^{i_6}),\\
g_4=&f(x) \Delta_{1,3|1,3}   = f(x)( x^{i_1+i_5}+x^{i_3}( x^{i_1}+x^{i_5}+x^{i_6} ),\\
g_5=&f(x) \Delta_{1,3|1,2}   = f(x)( x^{i_1}(x^{i_5}+x^{i_6}) +x^{i_2}( x^{i_1}+x^{i_5}+x^{i_6} ),\\
g_6=&f(x) \Delta_{1,3|1,4}   = f(x)( x^{i_1}(x^{i_4}+x^{i_6}) +x^{i_4}( x^{i_1}+x^{i_5}+x^{i_6} ).
\end{align*}
Then  the matrix $G$ is a polynomial generator matrix of the code.  Together with the following three components of total weight 36, 
{\small \begin{align*} (G_{short}M^\tr)[1] =& x^{63} + x^{53} + x^{41} + x^{37} + x^{36} + x^{25} + x^{22} + x^{20} + x^{18} + x^{7},\\ (G_{short}M^\tr) [2] =& x^{64} + x^{63 
   } + x^{59} + x^{51} + x^{50} + x^{45} + x^{43} + x^{42} + x^{39} + x^{38} + x^{25} + x^{18} + x^{8 
   } + x^{7}, \\(G_{short}M^\tr) [3] =& x^{50} + x^{45} + x^{43} + x^{42} + x^{40} + x^{39} + x^{37} + x^{28} + x^{23} + x^{22} + 
    x^{18} + x^{3}, \end{align*}}we obtain a codeword in $\ker H_{GC}$ of weight 88, which is likely the minimum distance of the code (the Magma program has $64\leq d\leq 88$). Therefore, we obtain a $[476, 72 , 64\leq d\leq 88]$ code. We denote this code as $C_2$ and will later provide simulation results for it in Section \ref{sec:sim}.     
    
   Note that we could have also continued to simplify $H_{short}$ as
 {\small \begin{align*}  &\sim\left[\begin{array}{c|ccc}1\!\!&\!\! x^{i_2-i_1}\!\!&\!\!x^{i_3-i_1}\!\!&\!\! x^{i_4-i_1}\\\hline0\!\!&\!\! x^{i_2}+ x^{i_5}+(x^{i_5}+x^{i_7})x^{i_2-i_1}\!\!&\!\!x^{i_5}+x^{i_7}+(x^{i_5}+x^{i_7})x^{i_3-i_1}\!\!&\!\!x^{i_4}+x^{i_7}+(x^{i_5}+x^{i_7})x^{i_4-i_1}\\0 \!\!&\!\!x^{i_5}+ x^{i_6}+(x^{i_1}+ x^{i_5}+ x^{i_6}) x^{i_2-i_1}\!\!&\!\! x^{i_5} +(x^{i_1}+ x^{i_5}+ x^{i_6})x^{i_3-i_1} \!\!&\!\!x^{i_4}+ x^{i_6} +(x^{i_1}+ x^{i_5}+ x^{i_6})x^{i_4-i_1}\end{array}\right]\end{align*}}
and thus consider the $[204, 72, 22\leq d\leq 34]$  code given by  $\ker(H_{short, 2})$ where  
{\small \begin{align*} &H_{short, 2}\defeq\\&\left[\begin{array}{ccc} x^{i_2}+ x^{i_5}+(x^{i_5}+x^{i_7})x^{i_2-i_1}\!\!&\!\!x^{i_5}+x^{i_7}+(x^{i_5}+x^{i_7})x^{i_3-i_1}\!\!&\!\!x^{i_4}+x^{i_7}+(x^{i_5}+x^{i_7})x^{i_4-i_1}\\
x^{i_5}+ x^{i_6}+(x^{i_1}+ x^{i_5}+ x^{i_6}) x^{i_2-i_1}\!\!&\!\!
 x^{i_5} +(x^{i_1}+ x^{i_5}+ x^{i_6})x^{i_3-i_1} \!\!&\!\!
 x^{i_4}+ x^{i_6} +(x^{i_1}+ x^{i_5}+ x^{i_6})x^{i_4-i_1}
 \end{array}\right] =\\&
\left[\begin{array}{ccc} x^{62} + x^{61} + x^{7} + x  &      x^{63} + x^{50} + x^{14} + x&     x^{58} + x^{45} + x^{44} + x^{14}\\
x^{62} + x^{61} + x^{46} + x^{39} + x  &      x^{50} + x^{49} + x^{27} + x  &   x^{46} + x^{45} + x^{22}\end{array}\right]. \end{align*} } The following is a generator matrix of weight 34 (which is the minimum distance of this code) 
 \begin{align*}G_{short,2} \defeq \begin{bmatrix} f_1&f_2&f_3\\\hline f_4&0& f_5\\  f_6& f_5&0\end{bmatrix},\end{align*}  where $f(x)(x^4+1)=x^{68}+1$ in $\Ftwo[x]$ and
 \begin{align*} f_1=&\Delta_{23} =x^{64} + x^{63} + x^{60} + x^{47} + x^{45} + x^{39} + x^{36} + x^{28} + x^{25} + x^{23} + x^{15} + x^{3},\\
  f_2= &\Delta_{13} =x^{60} + x^{59} + x^{51} + x^{47} + x^{45} + x^{40} + x^{39} + x^{38} + x^{37} + x^{36} + x^{22} + x^{15} + x^{8} + x^{7},\\
  f_3=& \Delta_{12} = x^{64} + x^{60} + x^{53} + x^{50} + x^{47} + x^{43} + x^{42} + x^{41} + x^{40} + x^{20} + x^{15} + x^{7},\\
 f_4= & f(x)(x^{i_4}+ x^{i_6} +(x^{i_1}+ x^{i_5}+ x^{i_6})x^{i_4-i_1}) = x^{65} + x^{61} + x^{57} + x^{53} + x^{49} + x^{45} + x^{41} + x^{37} \\+&x^{33} + x^{29} + x^{25} + x^{21} + x^{17} + x^{13} +   x^{9} + x^{5} + x,  \\ f_5=& f(x)(x^{i_5}+ x^{i_6}+(x^{i_1}+ x^{i_5}+ x^{i_6}) x^{i_2-i_1})= x^{67} + x^{63} + x^{59} + x^{55} + x^{51} + x^{47} + x^{43} + x^{39}\\ +& x^{35} + x^{31} + x^{27} + x^{23} + x^{19} + x^{15} +    x^{11} + x^{7} + x^3, \\
 f_6=& f(x)(x^{i_5} +(x^{i_1}+ x^{i_5}+ x^{i_6})x^{i_3-i_1})= x^{67} + x^{66} + x^{63} + x^{62} + x^{59} + x^{58} + x^{55} + x^{54} + x^{51} \\+& x^{50} + x^{47} + x^{46} + x^{43} + x^{42} + 
   x^{39} + x^{38} + x^{35} + x^{34} + x^{31} + x^{30} + x^{27} + x^{26} + x^{23} + x^{22} \\+& x^{19} + x^{18} + x^{15} + x^{14} +    x^{11} + x^{10} + x^{7} + x^{6} + x^{3} + x^{2}. \end{align*} 
  The code generated by the weight 38 matrix $\begin{bmatrix} f_1&f_2&f_3\end{bmatrix}$  has dimension 64 and minimum distance 38, and each row  $ \begin{bmatrix}
f_4&0& f_6\end{bmatrix}$ and $ \begin{bmatrix}  f_5& f_6&0\end{bmatrix}$ adds a rank of 4, giving a total rank of 72, while $ \begin{bmatrix}
f_4&0& f_6\end{bmatrix}$ has weight 34, which is the minimum distance of $\ker(H_{short,2})$. 
% We observe that the codewords can be viewed as $\vect{v}^\tr=(\vect{v}^\tr_1, \vect{v}^\tr_2)$ with $$\vect{v}_1\in \ker(\begin{matrix}   f_1&f_2 &f_3&f_4\end{matrix}), \vect{v}_2=M\vect{v}_1 ,$$
% so  $d_{min}(\ker(\begin{matrix}   f_1&f_2 &f_3&f_4\end{matrix}) )$ is a lower bound for the minimum distance of the code, and a generator matrix $G$ for $\ker(\begin{matrix}   f_1&f_2 &f_3&f_4\end{matrix}) $ is used to create the 
% generator matrix for the code as $\begin{bmatrix} G & GM^\tr\end{bmatrix}$. 
\pagebreak

\underline{Other matrices $M_1, M_2$} 

 Above,  we chose a certain matrix $h'_{GC}$ with the matrices $M_1$ and $M_2$, but this choice was arbitrary. % since it was the order of $x^{i_1}, \ldots, x^{i_7}$. 
 There are many other ways to choose  either of the two, and each might give a non-equivalent QC-GLDPC code. %We have tried three different others. 
 For example, the three versions below  
 {\small  \begin{align*} & \left[\begin{array}{c|c}  M_1^{\uparrow \{\begin{matrix}x^{i_1}& x^{i_2}& x^{i_3} & x^{i_4}\end{matrix}\}}&M_2^{\uparrow \{\begin{matrix} x^{i_5}&x^{i_6}&x^{i_7}\end{matrix}\}}\end{array}\right] =\begin{bmatrix} 
 1& x^{61}&x^{49}&0&x&0&0\\ 1& x^{61}&0& x^{44}&0&x^{46}&0\\1& 0&x^{49}& x^{44}&0&0&x^{14} \end{bmatrix},   \\& \begin{bmatrix} 
 x&0&0&1&  x^{61}&x^{49}&0\\0&x^{46}&0& 1&  x^{61}&0& x^{44} \\ 0&0&x^{14}& 1&0&x^{49}& x^{44}  \end{bmatrix}, \text{ and }  \begin{bmatrix} 0&x&0&0& 1& x^{61}&x^{49}\\  x^{44}&0&x^{46}&0& 1& x^{61}&0\\ x^{44}&0&0&x^{14}& 1&0&x^{49}\end{bmatrix},
\end{align*} }give the codes $[476, 71,  68\leq d\leq 80]$, $ [476, 69, 73\leq d\leq 86]$, and $[476, 72,  66\leq d\leq 88]$, respectively. \hfill $\Box$
 %\end{example}
 
 A short example demonstrating the procedure for larger constraint code weight is contained in Appendix \ref{app:ham15}.

\subsection{Generalizing after Observing Prelifting} 

For the $2\times 6$  protograph, it was shown in~\cite{lom18b} that the iterative decoding threshold was the closest to capacity when 75\% of the simple check nodes of the lifted $2N\times 6N$ scalar matrix were generalized with \eqref{shortHamming}. But this will, in most cases, break the QC structure and make the encoding less efficient. In order to  maintain the QC structure, and possibly create a stronger code (with better minimum distance), we will apply some of our previous techniques from~\cite{msc14} concerning double liftings (a ``pre-lift" and a ``final lift"). 
We recall Theorem 49 from~\cite{9762914}.

\begin{theorem}[\!\!\cite{9762914}, Theorem 49]  \label{thm:pre-lift} Every  QC-LDPC code with parity-check matrix $H$ of length $n_vN_1N_2$ is equivalent to a pre-lifted QC-LDPC code of with pre-lift size (first lifting factor) $N_1$ and circulant size (second lifting factor) $N_2$.% Similarly, every convolutional LDPC code with with parity-check matrix $H$ is  equivalent to a pre-lifted convolutional  code of any pre-lift size $N_1$.
\end{theorem} 
\begin{IEEEproof} 
%Since every cyclic code of size $N$ which is a multiple of $N_1$ is equivalent to an $N_1$-QC code of size $N_2$, \
We can transform each polynomial entry $g(x)= g_0(x^{N_1})+xg_1(x^{N_1})+\cdots + x^{N_1-1}g_{N_1-1} (x^{N_1})$ of $H$ into an $N_1\times N_1$ equivalent matrix, 
$$\begin{bmatrix} g(x) \end{bmatrix}= \begin{bmatrix} g_0(x)&xg_{N_1-1}(x)& xg_{N_1-2}(x) &\cdots & xg_{1}(x)\\ g_{1}(x)& g_0(x)&xg_{N_1-1}(x)&\cdots & xg_{N_1-2}(x)\\ \vdots&\vdots&\cdots &\vdots\\ g_{N_1-1}(x)& g_{N_1-2}(x) &g_{N_1-3}(x)&\cdots & g_0(x)\end{bmatrix}. $$
In the scalar matrix $H$ we can see this equivalence by performing a sequence of  column and row permutations (reordering of the columns and rows). We abuse the notation and use the equality sign between the two equivalent  representations (which result in  equivalent graphs). %$$\begin{bmatrix} g(x) \end{bmatrix}= \begin{bmatrix} g_0(x)&g_{N_1-1}(x)& g_{N_1-2}(x) &\ldots & g_{1}(x)\\ g_{1}(x)& g_0(x)&g_{N_1-1}(x)&\ldots & g_{N_1-2}(x)\\ \vdots&\vdots&\ldots &\vdots\\ g_{N_1-1}(x)& g_{N_1-2}(x) &g_{N_1-3}(x)&\ldots & g_0(x)\end{bmatrix}.$$ 
% So we replace each entry with an $N_1\times N_1$ matrix to obtain a matrix in a pre-lift form, with pre-lifting factor $N_1$. 
\end{IEEEproof} 
For example, if $N_1=2$ and $N=2N_2$, then the entries $x^{2a} =(x^2)^a$ and $x^{2a+1}=x (x^2)^a $ give the following transformations $$\begin{bmatrix} x^{2a} \end{bmatrix}= \begin{bmatrix} x^{a} &0\\0&x^a\end{bmatrix} \quad \text{ and } \quad \begin{bmatrix} x^{2a+1} \end{bmatrix}=\begin{bmatrix} 0& x^{a+1}\\x^a&0\end{bmatrix}. $$ 
\noindent  Similarly, if $N=3N_2$, the  equivalent code is $$\begin{bmatrix} x^{3a} \end{bmatrix}=\begin{bmatrix} x^{a} &0&0\\0&x^a&0\\0&0&x^a\end{bmatrix}, \begin{bmatrix} x^{3a+1} \end{bmatrix}=\begin{bmatrix} 0& 0&x^{a+1}\\x^a&0&0\\0&x^a&0\end{bmatrix}, \text{ and }\begin{bmatrix} x^{3a+2} \end{bmatrix}= \begin{bmatrix} 0& x^{a+1}&0\\0&0&x^{a+1}\\x^a&0&0\end{bmatrix},$$ with component matrices of size $N_2\times N_2$. 

\begin{corollary} Let $i_1=2j_1, \ldots, i_m=2j_m$ be the even exponents and $i_{m+1}=2k_1+1, \ldots, i_{n_v}=2k_n+1 $  the odd exponents, where $m+n=n_v$.  Then the matrix \begin{align*}
H\defeq \begin{bmatrix} 1&1&1&\cdots & 1\\ 1& x^{i_2}& x^{i_3}&\cdots&x^{i_{n_v}}\end{bmatrix}=\begin{bmatrix} 1&\ldots&1&1&\cdots & 1\\ x^{2j_1}&  \ldots& x^{2j_m}&x^{2k_1+1}& \cdots &x^{2k_n+1}\end{bmatrix}
\end{align*} is row and column  equivalent to the matrix 
\begin{align*}
\left[ \begin{array}{ccc|ccc|ccc|ccc}
1&\ldots&1&0&\ldots&0&1&\ldots&1 &0& \ldots&0\\
0&\ldots&0 &1&\ldots&1&0& \ldots&0&1&\ldots&1\\\hline 
x^{j_1}& \ldots& x^{j_m}&0&\ldots&0&0&\ldots&0&x^{k_1+1}& \ldots &x^{k_n+1}\\ 
0&\ldots&0&x^{j_1}& \ldots&x^{j_m}&x^{k_1}& \ldots&x^{k_n} &0&\ldots&0\end{array}\right].\end{align*}
\end{corollary} 
Let $h_{GC}=\begin{bmatrix} M_1&M_2 \end{bmatrix}$ be the matrix used to generalize $H$, with $M_1$ an $\ell\times m$ matrix and $M_2$ an $\ell\times n$ matrix, for some chosen $\ell>1$,  and $m+n=n_v$. By generalizing the first three constraints of this matrix using  $h_{GC}$ and the lifting factor $N/2$, we obtain the following parity-check matrix $H_{GC}$ for a GLDPC code
 \begin{align*}
&\left[\begin{array}{c|c|c|c}  
M_1&0&M_2&0\\
0&M_1&0&M_2\\
{M_1}^{\uparrow \{\begin{matrix} x^{j_1}& \cdots& x^{j_m}\end{matrix}\}}&0& 0&{M_2}^{\uparrow\{\begin{matrix} x^{k_1+1}& \cdots &x^{k_n+1}\end{matrix} \}}
\\ 0& \begin{matrix} x^{j_1}& \cdots&x^{j_m} \end{matrix}&\begin{matrix} x^{k_1}& \cdots&x^{k_n}\end{matrix}  &0\end{array}\right].
\end{align*}
%%%%%%%%%%%%%%%%%%%%%%%%%%%%%%%%%
Here we again use the notation %$M, I$ are the $3\times 3$ matrices such that  $h_{GC} =\left[\begin{array}{c|c} M&I\end{array}\right]$,  while  
$A^{\uparrow\{j_i\}}$ for  the generalized (expanded) matrix based on a matrix $A$, with the exponents reflecting the exponents in the generalized matrices. 

If  $\ell=n$ and $M_2=I_{n\times n}$, as we will consider in some of our cases, and if we denote   $M\defeq M_1$, for simplicity, then this matrix is equivalent to 
$$\left[\begin{array}{cc|cc}  
M&0&I&0\\
0&M&0&I\\\hline
{M}^{\uparrow \{\begin{matrix} x^{j_1}& \ldots& x^{j_m}\end{matrix}\}}&0& 0&I^{\uparrow\{\begin{matrix} x^{k_1+1}& \ldots &x^{k_n+1}\end{matrix} \}}
\\ 0& \begin{matrix} x^{j_1}& \ldots&x^{j_m} \end{matrix}&\begin{matrix} x^{k_1}& \ldots&x^{k_n}\end{matrix}  &0\end{array}\right]\sim 
$$

 \begin{align*}
\\&H_{GC}\defeq\left[\begin{array}{cc|cc}  
M&0 &I&0\\
0&M&0&I\\\hline{M}^{\uparrow \{\begin{matrix} x^{j_1}& \ldots& x^{j_m}\end{matrix}\}}&I^{\uparrow\{\begin{matrix} x^{k_1+1}& \ldots&x^{k_n+1}\end{matrix} \}}M &0& 0\\
\begin{bmatrix} x^{k_1}& \ldots &x^{k_n} \end{bmatrix}M &\begin{bmatrix} x^{j_1}& \ldots& x^{j_m} \end{bmatrix}&0&0\end{array}\right]\sim 
\left[\begin{array}{c|c} \diag(M,M) & I_{2n\times 2n}\\\hline   H_{1,short}& 0_{(n+1)\times n}\end{array}\right]
,\end{align*}%$$
where $$H_{1,short}\defeq \left[\begin{array}{cc}{M}^{\uparrow \{\begin{matrix} x^{j_1}& \ldots& x^{j_m}\end{matrix}\}}&I^{\uparrow\{\begin{matrix} x^{k_1+1}& \ldots&x^{k_n+1}\end{matrix} \}}M \\
\begin{bmatrix} x^{k_1}& \ldots &x^{k_n} \end{bmatrix}M &\begin{bmatrix} x^{j_1}& \ldots& x^{j_m} \end{bmatrix}\end{array}\right] \sim$$ $$ \left[\begin{array}{cc}\left(I^{\uparrow\{\begin{matrix} x^{k_1+1}& \ldots&x^{k_n+1}\end{matrix} \}}\right)^{-1}{M}^{\uparrow \{\begin{matrix} x^{j_1}& \ldots& x^{j_m}\end{matrix}\}}&M \\
\begin{bmatrix} x^{k_1}& \ldots &x^{k_n} \end{bmatrix}M &\begin{bmatrix} x^{j_1}& \ldots& x^{j_m} \end{bmatrix}\end{array}\right].$$
%$$\sim \left[\begin{array}{cc}
%\begin{bmatrix} x^{k_1}& \ldots &x^{k_n} \end{bmatrix}M &\begin{bmatrix} x^{j_1}& \ldots& x^{j_m} \end{bmatrix}\\ \left(I^{\uparrow\{\begin{matrix} x^{k_1+1}& \ldots&x^{k_n+1}\end{matrix} \}}\right)^{-1}{M}^{\uparrow \{\begin{matrix} x^{j_1}& \ldots& x^{j_m}\end{matrix}\}}&M \end{array}\right]
%$$
Following Theorem~\ref{thm-directly}, a generator matrix for $G$ for $\ker (H_{GC})$  is 
 $$\begin{bmatrix} G_{1,short} &G_{1, short}\diag(M^\tr,M^\tr)\end{bmatrix},$$ where $G_{1, short}$ is a generator matrix for $\ker (H_{1, short})$.
 
We will show how to do this  for the example of $n_v=6$ (and $n_v=7$ in Appendix \ref{app:pre2}). In particular, we will show how to simplify $H_{1, short}$ in the same manner that we simplified $H_{GC}$. 

\noindent{\bf Example \ref{example1-GLDPC} (cont.).} %\begin{example}
Consider again $n_v=6$ and $[i_1,i_2,i_3,i_4,i_5,i_6]= [1, x^{54}, x^{66},x^{71}, x^{55} ,x^{69}]$ from earlier. 
 First, we increase the lifting exponent to $N=90$, which is the first even exponent for which the girth of the $2\times 6$ constraint matrix is 12. Then, we generalize the first constraints using $h_{GC}\defeq \begin{bmatrix} M&I\end{bmatrix}$ from the examples before and  obtain $H_{1, short}$ (below). 

The code $\ker(H_{1, short})$ is a  $[270, 91, 27]$ code, its minimum distance is a lower bound on the distance of the original code. We will further simplify this matrix by row reducing $H_{1, short}$ as 

%\begin{figure*}
\begin{align*} H_{1, short}\sim &
\left[\begin{array}{ccc|ccc} x^{35}+x^{27}& x^{35}+x^{34}& x^{27}+x^{34}&1& x^{27}&x^{33}\\\hline x^{-36}& x^{27-36}&0& 1& 1 &0\\x^{-28}&0&x^{33-28}&1&0& 1\\0&x^{27-35}&x^{33-35}&0& 1& 1
\end{array}\right]\sim\\
&\left[\begin{array}{ccc|ccc} x^{35}+x^{27}& x^{35}+x^{34}& x^{27}+x^{34}&1& x^{27}&x^{33}\\x^{-36}& x^{27-36}&0& 1& 1 &0\\x^{-28}&0&x^{33-28}&1&0& 1\\\hline x^{-36}+x^{-28}&x^{27}(x^{-35}+x^{-36})&x^{33}(x^{-35}+x^{-28}) &0& 0& 0
\end{array}\right]=\\& \left[\begin{array}{c|c} A &B\\\hline \begin{matrix}x^{-36}+x^{-28}&x^{27}(x^{-35}+x^{-36})&x^{33}(x^{-35}+x^{-28}) \end{matrix} &0\end{array}\right]
\sim\\& \left[\begin{array}{c|c} B^{-1}A &I \\\hline\begin{matrix}x^{-36}+x^{-28}&x^{27}(x^{-35}+x^{-36})&x^{33}(x^{-35}+x^{-28}) \end{matrix} &0\end{array}\right], \end{align*}
where
$B^{-1}= (1+x^{27}+x^{33})^{-1} \begin{bmatrix} 1&x^{27}&x^{33}\\ 1&1+x^{33}&x^{33}\\ 1&x^{27}&1+x^{27}\end{bmatrix}, $  from which, 
{\small $$B^{-1}A=(1+x^{27}+x^{33})^{-1} \begin{bmatrix} x^{36} + x^{35} + x^{27} + x^5  &x^{35} + x^{34} + x^{18}&      x^{38} + x^{34} + x^{27}\\
  x^{42} + x^{35} + x^{27} + x^9 + x^5 &       x^{36} + x^{35} + x^{34} + x^{24}&      x^{38} + x^{34} + x^{27}\\
x^{44} + x^{36} + x^{35} + x^{27} + x^{17}&      x^{35} + x^{34} + x^{18}&x^{34} + x^{32} + x^{27} + x^5
 \end{bmatrix}. $$}We do not need to substitute $ (1+x^{27}+x^{33})^{-1} =x^{30} + x^{27} + x^{18} + x^{15} + x^{12} + x^9 + x^6 + x^3 + 1.$  % \\ &(1+x^{27}+x^{33})^{-1} =x^{30} + x^{27} + x^{18} + x^{15} + x^{12} + x^9 + x^6 + x^3 + 1
%Therefore, 
%\begin{align*} &\left[\begin{array}{c|c}  H_{1, short}& 0\\\hline\begin{matrix} M&0\end{matrix}& \begin{matrix} I&0\end{matrix}\\\begin{matrix} 0&M\end{matrix}& \begin{matrix} 0&I\end{matrix}\end{array}\right]=\left[\begin{array}{c|c||c|c}  B^{-1}A &I&0&0\\\begin{matrix}x^{-36}+x^{-28}&x^{27}(x^{-35}+x^{-36})&x^{33}(x^{-35}+x^{-28}) \end{matrix} &0&0&0\\\hline
%M&0 & \ I&0 \\ 0&M&  0&I\end{array}\right] \end{align*}
%So $\ker(H_{1, short})$ consists of vectors $(\vect{v}^\tr_1,\vect{v}^\tr_3)^\tr$ such that $${\small\vect{v}_1 \in \ker\begin{bmatrix}x^{-36}+x^{-28}&x^{27}(x^{-35}+x^{-36})&x^{33}(x^{-35}+x^{-28}) \end{bmatrix}}, \quad \vect{v}_3 =B^{-1}A\vect{v}_1, $$ while  $(\vect{v}^\tr_2,\vect{v}^\tr_4)^\tr= \left(\vect{v}^\tr_1 M^\tr, \vect{v}^\tr_3 M^\tr\right)^\tr$. Therefore, $$(\vect{v}^\tr_1,\vect{v}^\tr_2\vect{v}^\tr_3,\vect{v}^\tr_4)= \vect{v}^\tr_1\begin{bmatrix} I&M^\tr & (B^{-1}A)^\tr & (MB^{-1}A)^\tr\end{bmatrix}.$$
Following Theorem~\ref{thm-directly},  if  $G_{2, short}$ is a polynomial generator matrix for the code $\ker(H_{2,short})$, where $$H_{2,short}\defeq\begin{bmatrix}x^{-36}+x^{-28}&x^{27}(x^{-35}+x^{-36})&x^{33}(x^{-35}+x^{-28}) \end{bmatrix}, $$ then $G_{1,short}\defeq \begin{bmatrix}G_{2, short} & G_{2, short}(B^{-1}A)^\tr \end{bmatrix}$
is a generator matrix for $\ker(H_{1, short})$.  %or, m
%Multiplying by $B^\tr$ to the right, we obtain a generator matrix for an equivalent code
%$G'_{1,short}\defeq\begin{bmatrix}G_{2, short}B^\tr& G_{2, short}A^\tr \end{bmatrix}.$ \\
Therefore, 
\begin{align*}G\defeq & \begin{bmatrix} G_{1,short} &G_{1, short}\diag(M^\tr,M^\tr)\end{bmatrix}=\\
&\begin{bmatrix}G_{2, short} & G_{2, short}(B^{-1}A)^\tr &G_{2, short} M& G_{2, short}(B^{-1}A)^\tr M^\tr\end{bmatrix},\end{align*} and,  the sparser, $$G'\defeq (1+x^{12}+x^{18})\begin{bmatrix}G_{2, short} & G_{2, short}(B^{-1}A)^\tr &G_{2, short} M^\tr& G_{2, short}(B^{-1}A)^\tr M^\tr\end{bmatrix},$$
are generator matrices for the large  code.\footnote{A generator matrix in this form allows for a very simple encoding, starting from an encoder $G_2$ which can be easily generated, either with Theorem~\ref{case1} or using the Magma program. The polynomial $(1+x^{12}+x^{18})$ does not  need to be multiplied through because polynomial multiplication is easily implemented in circuitry. Decoding could also be devised based on the dependence of the  vectors $\vect{v}_2,\vect{v}_3,$ and $\vect{v}_{4}$ on $\vect{v}_1$. } 

%
%Alternatively, $$G''\defeq \begin{bmatrix} G'_{1,short} &G'_{1, short}\diag(M^\tr,M^\tr)\end{bmatrix}=
%\begin{bmatrix}G_{2, short}B^\tr & G_{2, short}A^\tr &G_{2, short}B^\tr M^\tr& G_{2, short}A^\tr M^\tr\end{bmatrix},$$ is a generator matrix for an equivalent code. 
%%$\vect{v}^\tr_1 =\vect{u}^\tr G_{2}$, for some vector $\vect{u}^\tr $,  and 
%% \begin{align*} &(\vect{v}^\tr_1,\vect{v}^\tr_2\vect{v}^\tr_3,\vect{v}^\tr_4)= \vect{u}^\tr\begin{bmatrix}G_{2} &G_{2}M^\tr & G_{2}(B^{-1}A)^\tr &G_{2} (MB^{-1} A)^\tr\end{bmatrix}, \end{align*}
% $$\begin{bmatrix}G_{2, short} &G_{2, short}M^\tr & G_{2, short}(B^{-1}A)^\tr &G_{2, short} (M B^{-1}A)^\tr\end{bmatrix},$$  and,  the sparser,  $$G\defeq (1+x^{12}+x^{18})\begin{bmatrix}G_2 &G_2M^\tr & G_2(B^{-1}A)^\tr &G_2 (M B^{-1}A)^\tr\end{bmatrix},$$
% $\begin{bmatrix}G_{2, short}B^\tr& G_{2, short}A^\tr \end{bmatrix},$ is a generator matrix for $\ker{H_{1, short}}.$
% or, alternatively,  are generator matrices for the original code,   $\ker(H'_{GC})$.
%i.e., 
%$$H'\begin{bmatrix}G_{2} &G_{2}M^\tr & G_{2}(B^{-1}A)^\tr &G_{2} (M B^{-1}A)^\tr\end{bmatrix}^\tr=0.$$

For example, using Theorem~\ref{case1}, we obtain the following polynomial generator matrix $G_{2, short}$ for the rate $91/135$ code $\ker(H_{2,short})$, % of minimum distance 3, where $$H_{2,short}=\begin{bmatrix}x^{-36}+x^{-28}&x^{27}(x^{-35}+x^{-36})&x^{33}(x^{-35}+x^{-28}) \end{bmatrix} :$$ 
$$G_{2, short} =\begin{bmatrix}x^{-27}(x^{35}+x^{36}) &x^{36}+x^{28}&0 \\x^{-33}(x^{35}+x^{28})& 0&x^{36}+x^{28} \\ f&0&0 \\ 0&f&0 \\ 0&0&f\end{bmatrix}, $$ where $f(x)$ is defined such that $f(x)(1+x)=1+x^{45}$.
%i.e., 
%  $${\small \begin{bmatrix}x^{-36}+x^{-28}&x^{27}(x^{-35}+x^{-36})&x^{33}(x^{-35}+x^{-28}) \end{bmatrix}} G_{2, short}^\tr=0.$$
%  $G_{2}$ generates a $91/135$ rate code, so the dimension of this shortened version $\ker{H_{2,short}}$ is 91, like $\ker(H_{1, short})$ and $\ker(H')$. 
The matrix $G_{2, short}$, however, does not contain as a row a codeword of the smallest weight 3. We can easily find such a codeword,  due to the simple structure of the matrix $H_{2,short}$.  We can visually  check that $ u(x)\begin{bmatrix} 1 & x^{27} & x^{33}\end{bmatrix} $ %= (1, x^{12}, x^{18})$
 is a codeword, for every polynomial $u(x)$. %Also, we observe that the linear combination  of the two entries of $G_2$,  $x^{27}G_2(1,1) +x^{33} G_2(1,1) =x^{28}+x^{36} = G_2(1,2)=G_2(2,3)$
Then a codeword in $\ker(H_{GC})$ is 
\begin{align*}\vect{v}=& u(x)\begin{bmatrix} 1 & x^{27} & x^{33}\end{bmatrix}  \begin{bmatrix} I & (B^{-1}A)^\tr &M^\tr& (MB^{-1}A)^\tr\end{bmatrix} \\
=&u(x) \left[\begin{array}{ccccccccccccccc} 1& %\quad  %1+x^{-27}+x^{-33}= 1+x^{12}+x^{18}, 
%v_2=
x^{27}& %\quad v_3= 
x^{33}&&  %\begin{matrix}v_1+v_2& v_1+v_3& v_2+v_3\end{matrix} =u(x)\
1+x^{27}& 1+ x^{33}& x^{27}+ x^{33}&&
% v_4 =v_1+v_2 =(1+x^{27})u(x),\quad 
%v_5= v_1+v_3= (1+ x^{33})u(x), \quad  v_6=v_2+v_3= (x^{27}+ x^{33})u(x),\\
v(x)&v(x)&v(x)&&0&0&0 \end{array}\right],
\end{align*}
where $v(x)= (1+x^{27}+x^{33})^{-1}( x^{44} + x^{38} + x^{37} + x^{18} + x^{10} + x^{6}). $%v_7= v_8= v_9= u(x)(1+x^{27}+x^{33})^{-1}( x^{44} + x^{38} + x^{37} + x^{18} + x^{10} + x^{6}), \\
%&\vect{v}^\tr_4 =% v_{10}= v_{11}= v_{12}=0. 

%We observe that any codeword $\vect{v}$ must satisfy $v_4+v_5+v_6= 0$ and $v_{10}+ v_{11}+v_{12}=0$ due to the multiplying by the matrix $M^\tr$.

Taking $u(x)=(1+x^{27}+x^{33})$, we obtain a codeword with $ v(x) = x^{44} + x^{38} + x^{37} + x^{18} + x^{10} + x^{6}$ of weight 6, giving a total vector $\vect{v}$ of weight 39, equal to the minimum distance. The part $( \vect{v}_1, \vect{v}_2 )$ is a codeword of smallest weight 27 in $\ker(H_{1, short})$, and the part $\vect{v}_1$ is a codeword in $\ker(H_{2, short})$ of weight 9. 
The resulting code is a $[540, 91, 39]$ code. 
  We can similarly obtain other vectors of small weight, starting from a different vector of weight 3, and thus obtain a polynomial matrix $G$ of the smallest possible weight.\footnote{Note that we can also find $\vect{v}$ directly from the matrix $G_2$ above. 
 Since $\gcd(x^{28}+x^{36}, f(x))=1$ in $\Ftwo[x]$, there exists 
$u(x)$ and $v(x)$ such that $u(x)(x^{28}+x^{36}) + v(x)f(x)=1, $  in $\Ftwo[x], $ where 
$u(x)= x^{42} + x^{39} + x^{38} + x^{36} + x^{35} + x^{33} + x^{32} + x^{29} + x^{26} + x^{23} + x^{22} + x^{20} + x^{19} + x^{17} + x^{16} + x^{13} + x^{10} + x^9 + x^7 + x^6 + x^4 + x^3 + 1.$
Therefore, in $ \Ftwo[x]/(x^{45}+1), $ we have that  $u(x)(x^{28}+x^{36}) + f(x)=1.$ If we take  $\vect{u}\defeq \left(x^{27}u(x), x^{33}u(x), 1,1,1\right),$
then 
$\vect{v}\defeq \vect{u} G$ is the  codeword of  weight 39 that we obtain directly in the text above,   with $u(x)=(1+x^{27}+x^{33})$. } 
\hfill $\Box$% \end{example}

\pagebreak
We can also  use different matrices for the generalized constraint nodes. 

%\begin{example} 
\noindent{\bf Example \ref{example1-GLDPC} (cont.).} For example, we generalized the first two nodes using  $
h_{GC, 0}\defeq \begin{bmatrix} M&I \end{bmatrix} %\defeq \begin{bmatrix} 1&1&0&1&0&0\\1&0&1&0&1&0\\ 0&1&1&0&0&1\end{bmatrix} 
$
%\vspace{-.15in}
and the third  using $h_{GC,1}\defeq \begin{bmatrix} I&M\end{bmatrix} $ to obtain a matrix $H_{GC}$. 
Applying the same technique, and Theorem~\ref{thm-directly}, we obtain that a generator matrix for the generalized code $\ker(H_{GC})$ is $$\begin{bmatrix} G_{1,short} &G_{1, short}\diag(M^\tr,M^\tr)\end{bmatrix},$$ where $G_{1, short}$ is a generator matrix for the shorter code with parity-check matrix 

 \begin{align*} &H_{1, short} \defeq 
\left[\begin{array}{c|c} I&\left(I^{\uparrow\{1,27,33\}}\right)^{-1} M^{\uparrow\{36,28, 35\}}M\\\begin{bmatrix}  x^{35}&x^{27}&x^{34} \end{bmatrix}M & \begin{bmatrix} 1&x^{27} & x^{33} \end{bmatrix} \end{array}\right]\sim
%\begin{bmatrix} 1& 0&0&x^{36}+x^{28}& x^{36}&x^{28}\\0&x^{27}&0&x^{36}&x^{36}+x^{35}& x^{35}\\0&0&x^{33}& x^{28}& x^{35}&x^{28}+ x^{35} \\x^{35}+x^{27}& x^{35}+x^{34}& x^{27}+x^{34}&1&x^{27}&x^{33}\end{bmatrix}\sim\\ 
%&\left[\begin{array}{ccc|ccc} 1& 0&0&x^{36}+x^{28}& x^{36}&x^{28}\\0&1&0&x^{9}&x^{9}+x^{8}& x^{8}\\0&0&1& x^{33}& x^{2}&x^{33}+ x^{2} \\\hline0& 0& 0&f_1& f_2&f_3\end{array}\right]=
\left[\begin{array}{c|c}I&A\\\hline0& \begin{matrix} f_1& f_2&f_3\end{matrix}\end{array}\right],\\&
\left\{\begin{matrix} f_1&=1+x^5+x^6+x^{17}+x^{22}+x^{29}+x^{33}  \\f_2&=x^4+x^6+x^{25}+x^{27}+x^{29}+x^{33}+x^{36} \\f_3&=x^4+x^5+x^{17}+x^{22}+x^{25}+x^{33}+x^{36}\end{matrix} \right.,    A\defeq \left[\begin{array}{ccc} x^{36}+x^{28}& x^{36}&x^{28}\\x^{9}&x^{9}+x^{8}& x^{8}\\ x^{33}& x^{2}&x^{33}+ x^{2} \end{array}\right].
 \end{align*}%  y^35+y^27, y^35+y^34, y^27+y^34,1, y^27,y^33,1, y^27,0, y^36, y^36,0,1,0,y^33,y^28,0, y^28,0,y^27,y^33,0, y^35, y^35

The code given by $H_{1, short}$  is a $[270, 90, 30]$ code. 
Applying Theorem~\ref{thm-directly}, if $G_{2,  short}$ is a polynomial generator matrix for the code $\ker(H_{2,short})$, where $H_{2,short}=\begin{bmatrix} f_1& f_2&f_3\end{bmatrix}, $ then 
$\begin{bmatrix} G_{2,short}A^\tr &G_{2, short}\end{bmatrix}$ is a polynomial generator matrix for the code $\ker(H_{1,short})$, 
%
%$\vect{v}^\tr_3 =\vect{u}^\tr G_{2}$, for some vector $\vect{u}^\tr $,  and 
% \begin{align*} &(\vect{v}^\tr_1,\vect{v}^\tr_2\vect{v}^\tr_3,\vect{v}^\tr_4)= \vect{u}^\tr\begin{bmatrix}G_{2}A^\tr &G_{2}(MA)^\tr  &G_{2} & G_{2}M^\tr \end{bmatrix}, \end{align*}
and thus, $$\begin{bmatrix}G_{2, short}A^\tr   &G_{2, short}&G_{2, short}A^\tr M^\tr& G_{2, short}M^\tr  \end{bmatrix}$$ 
% and,  possibly sparser,  $$G\defeq (1+x^{12}+x^{18})\begin{bmatrix}G_2 &G_2M^\tr & G_2(B^{-1}A)^\tr &G_2 (M B^{-1}A)^\tr\end{bmatrix},$$ are 
 is a generator matrix for the original code,   $\ker(H_{GC})$.  
 
 For example, using Theorem~\ref{case1}, we obtain the following polynomial generator matrix $G_{2, short}$ for the %rate $91/135$ 
 code $\ker{H_{2,short}} $ of minimum distance 9 as
$$G_{2} =\begin{bmatrix}f_2^\tr&f_1^\tr&0 \\f_3^\tr&0&f_1^\tr \end{bmatrix}. $$
The rows of $G_2$ can be used to find codewords in the original code, but they do not lead to  codewords of the smallest weight 40.
A codeword of weight 40 can be obtained by using the structure of the codewords as follows. (We obtained them by reverse engineering.)
Since $ f_1+f_2+f_3=1+x^{27}+x^{33}$ is invertible, we can take $\vect{v}_2= (v,v, v+x^{a})$ (because, when we multiply them with M, they will have as close to 0 weight as possible), such that

\pagebreak
  $$H_{2, short} \vect{v}^\tr_2 =f_1v^\tr(x)+f_2v^\tr(x)+f_3v^\tr(x)+f_3x^{-a}= (f_1+f_2+f_3)v^\tr(x)+
x^{-a} f_3=0.$$

Taking    $v^\tr(x)= x^{-a}(f_1+f_2+f_3)^{-1}f_3$, we obtain a codeword for any $a$.  We can also choose 
 $\vect{v}_2= (v+x^{a},v, v)$ or  $\vect{v}_2= (v,v+x^{a}, v)$, 
and  choose $i$ such that $(f_1+f_2+f_3)^{-1}f_i$ has the smallest weight, for $i=1,2,3$. In this case, it is for $i=3$, so we take \begin{align*}v^\tr(x)= &x^{-a}  (f_1+f_2+f_3)^{-1}f_3 = \\&x^{-a}(x^{36} + x^{32} + x^{22} + x^{21} + x^{16} + x^{10} + x^5 + x^4 + 1), \end{align*} because
\begin{align*}(f_1+f_2+f_3)^{-1}=x^{36 }+ &x^{34 }+ x^{33 }+ x^{31 }+ x^{27 }+ x^{26 }+ x^{25 }+ x^{23 }+ x^{21 }+ \\&x^{18 }+ x^{15 }+ x^{14 }+ x^{13 }+ x^{12 }+ x^{9 }+ x^8 + x^2 + x + 1.\end{align*}For example, $a=1$ gives $\vect{v}_2= (v,v, v+1)$, with $$v^\tr= x^{36} + x^{32} + x^{22} + x^{21} +x^{16} + x^{10} + x^5 + x^4 + 1.$$ Then, the weight of  $\vect{v}_4=\vect{v}_2M^\tr$  is 2, for any $a$, and any position of $x^{-a}$,  having a component equal to 0 and two equal to $x^{-a}$. Then  $\vect{v}_1=\vect{v}_2A^\tr$ is $x^a $ times one of  columns of $A$, depending on the component in which $x^{-a}$ is placed in $\vect{v}^\tr_2$, so it has weight 4, for any $a$. Finally, the weight of $\vect{v}_3=\vect{v}_1M^\tr$ has maximum weight $ {\rm wt}(\vect{v}_3)=3+2+3=8$, giving a total of $3 {\rm wt}(v)-1+2+8+4=13+3{\rm wt}(v)$.
 For $a=20$ and $v$ above, we get   $13+27=40$.\hfill $\Box$ 
% \end{example}
An additional example of this procedure, continuing Example \ref{example2-GLDPC} is given in Appendix \ref{app:pre2}.  

\subsection{Numerical Results}\label{sec:sim}
\begin{figure}[h]
  \centering
  \includegraphics[scale=0.5]{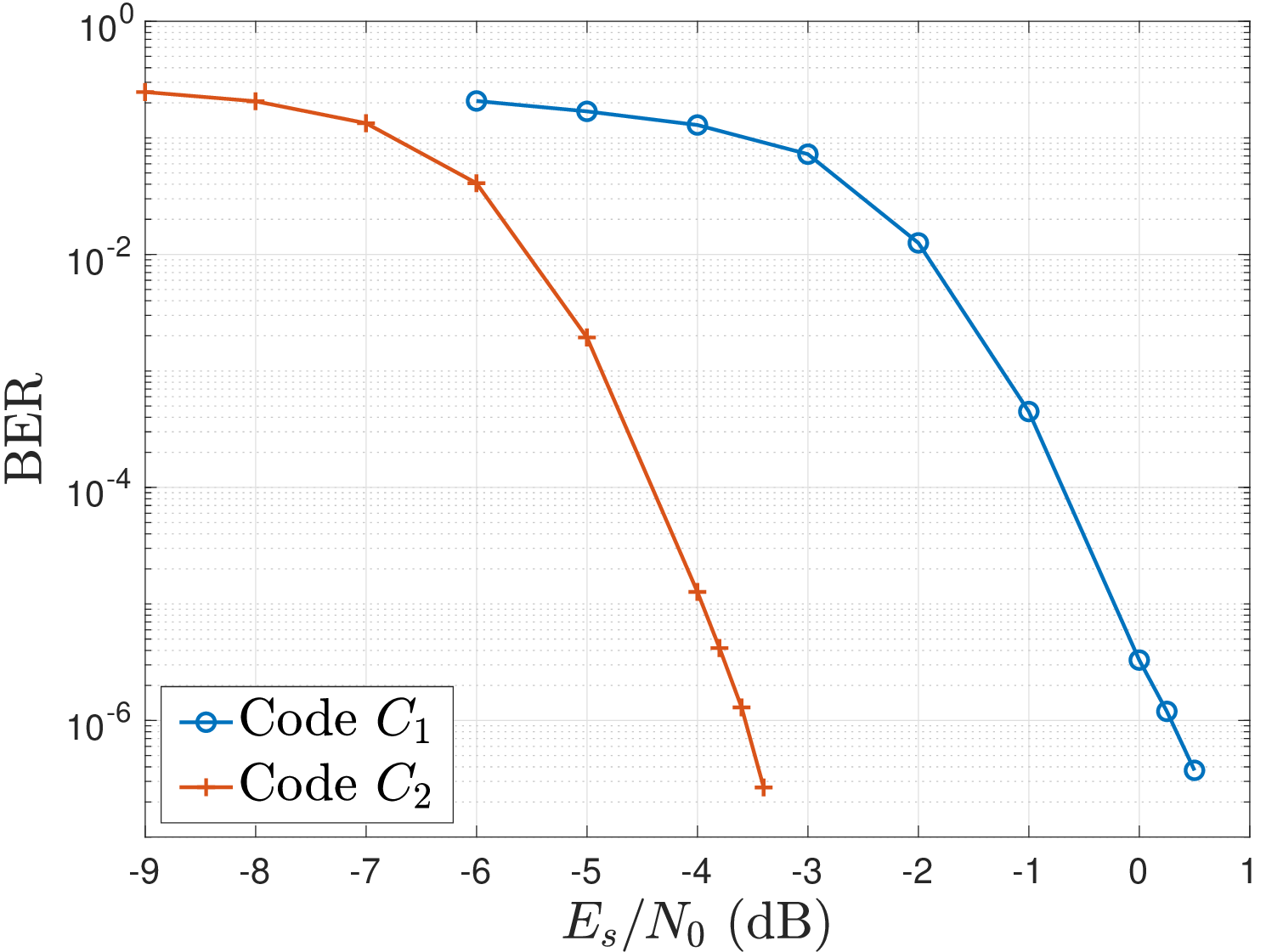}
  \caption{Computer simulation results obtained for the partially generalized $[476, 204, 16]$ QC-GLDPC code $C_1$ and fully generalized $[476, 72 , 64\leq d\leq 88]$ QC-GLDPC code $C_2$ from Example 5.}
  \label{fig:sim}
\end{figure}

Fig.~\ref{fig:sim} presents some simulation results of two QC-GLDPC codes from Example 5, one partially generalized and one fully generalized. Specifically, code $C_1$ is a $[476, 204, 16]$ partially generalized QC-GLDPC code and code $C_2$ is a $[476, 72 , 64\leq d\leq 88]$ fully generalized QC-GLDPC code. Simulations were performed using belief propagation (BP) decoding on the non-generalized parity-check matrix with a maximum number of $100$ iterations, where the SPC constraints are updated as the conventional sum-product algorithm (SPA), but the soft-outputs from the generalized constraints are decoded using the Bahl-Cocke-Jelinek-Raviv (BCJR) algorithm (see \cite{lrc08} for details).\footnote{We display the results in terms of $E_s/N_0$ (dB) since the codes have different rates.} Log-likelihood ratios are clipped at $\pm 20$ and for the BCJR algorithm we set the forward, backward, and branch metrics to have a threshold of $\pm 2.5\times 10^4$. As expected from the high girth and good minimum distance, the codes display a relatively steep waterfall and no indication of an error floor down to a bit error rate (BER) of $10^{-6}$.

%%%%%%%%%%%%%%%%%%%%%%%%%%%%%%%%%%%%%%%%%%%%%%%%%%%%
\section{Concluding Remarks}\label{sec:conc}
%%%%%%%%%%%%%%%%%%%%%%%%%%%%%%%%%%%%%%%%%%%%%%%%%%%%
This paper shows how to obtain a generator matrix for a QC-GLDPC code using minors of the polynomial parity-check matrix. The approach can be applied to fully generalized matrices, or partially generalized (with mixed constraint nodes) in order to find better performance/rate trade-offs. The resulting matrices can be presented in several forms that may facilitate efficient encoder implementation as well as minimum distance analysis. These forms allow us to display a variety of codewords and possibly, those of weight equal to the minimum distance. %{question} NOTE: WE NEED TO SHOW THIS IF WE CLAIM IT: 
%Moreover, the approach was shown to provide a formula for the dimension of any QC-LDPC code based on the minors of its polynomial parity-check matrix. 
We also  demonstrated that the code parameters can be improved by applying a double graph-lifting procedure that does not affect the ability to obtain a polynomial generator matrix.
%%%%%%%%%%%%%%%%%%%%%%%%%%%%%%%%%%%%%%%%%%%%%%%%%%%    
%%%%%%%%%%%%%%%%%%%%%%%%%%%%%%%%%%%%%%%%%%%%%%%%%%%
\section*{Acknowledgements}
The authors would like to sincerely thank Salman Habib for help with the computer simulation results.%, and fellowships from  GFSD and Kinesis-Fern\'andez Richards.
\balance
\bibliographystyle{IEEEtran}
\bibliography{newprefs.bib}

 \appendices
 %%%%%%%%%%%%%%%%%%%%%%%%%%%%%%%%%%%%%%%%%%%%%%%%
  %%%%%%%%%%%%%%%%%%%%%%%%%%%%%%%%%%%%%%%%%%%%%%%%
 \section{QC-GLDPC Codes with High Degree Constraint Nodes}\label{app:ham15}
 %%%%%%%%%%%%%%%%%%%%%%%%%%%%%%%%%%%%%%%%%%%%%%%%
%%%%%%%%%%%%%%%%%%%%%%%%%%%%%%%%%%%%%%%%%%%%%%%%
 \begin{example}\label{first15} We first consider the partial generalization of \eqref{matrix2byn} with $n_v=15$. Here, we suppose that the component codes are again Hamming codes, with  $h_{GC}\defeq\begin{bmatrix} M&I \end{bmatrix}$ being a $4\times 15$ parity-check matrix for the Hamming $[15,11,3]$ code. 
%\vspace{-5pt}
%{ \begin{align*}% &h_{GC,0}\defeq\begin{bmatrix} M_1&M_2 \end{bmatrix} =\\
%h_{GC}\defeq\begin{bmatrix} I&M \end{bmatrix}=&\left[\begin{array}{cccccccccccccccc} 
%1&0&0&0&&1&0&1&1& 1&0&1&0&1&0&1\\
%0&1&0&0&&0&1&1&1& 0&1&1&0&0&1&1 \\ 
%0&0&1&0&&1&1&1&0& 0&0&0&1&1&1&1\\  
%0&0&0&1&&0&0&0&0& 1&1&1&1&1&1&1
%\end{array}\right] \end{align*}.}
%\vspace{-1pt}  and 
%$h_{GC,1}\defeq \begin{bmatrix} M_2&M_1 \end{bmatrix}$ following \cite{lrc08}.
%We take $N=31$  and the exponents $[i_2, i_3, \ldots , i_{15}]=$  $ [6,23, 3 , 7,21,1, 18,9,27,19,28,16,17,20]$,  for  which $H$ has girth 8 to 
We obtain $H_{GC}$ given by
%\vspace{-5pt}
\begin{align*}
%\label{largech6}
H_{GC}
=\arraycolsep=2.54pt\left[\begin{array}{cccccccccccc|cccc} %0&1&2&3&4&5&6&7&&8 &9& 10&11&12&13&14\\\hline 
1&x^{i_2}&x^{i_3}&x^{i_4}&x^{i_5}&x^{i_6}&x^{i_7}&x^{i_8}&x^{i_9}&x^{i_{10}}&x^{i_{11}}&&x^{i_{12}}&x^{i_{13}}&x^{i_{14}}  &x^{i_{15}}\\\hline  
1&0&1&1& 1&0&1&0&1&0&1&&1&0&0&0\\
0&1&1&1& 0&1&1&0&0& 1&1&& 0&1&0&0\\ 
1&1&1&0&  0&0&0&1& 1&1&1&& 0&0&1&0\\  
0&0&0&0& 1&1&1&1&1&1&1&&0&0&0&1\end{array}\right]
\sim \left[\begin{array}{c|c} H_{short}&\vect{0} \\\hline M&I\end{array}\right] ,\end{align*}
where $ H_{short}\defeq \begin{bmatrix}   f_1&\cdots  &f_{11}\end{bmatrix}\defeq \begin{bmatrix} 1& x^{i_2}&\cdots& x^{i_{11}} \end{bmatrix} +\begin{bmatrix}  x^{i_{12}}&\cdots&x^{i_{15}}\end{bmatrix} M.$
%\eqref{largech6}. 
We apply  Theorem~\ref{thm-directly}  to find a generator matrix. The code has an approximate rate of  $10/15=2/3$. For example, for $N=376$, the following choice gives a girth 12 code. 
\begin{align*}&\left[\begin{array}{ccccccccccccccc} 1&x^{i_2}&x^{i_3}&x^{i_4}&x^{i_5}&x^{i_6}&x^{i_7}&x^{i_8}&x^{i_9}&x^{i_{10}}&x^{i_{11}}&x^{i_{12}}&x^{i_{13}}&x^{i_{14}}  &x^{i_{15}}\end{array}\right]= \\&
\left[\begin{array}{ccccccccccccccc}1& x^{14}& x^{ 24}& x^{44}& x^{46}& x^{180}&x^{ -100}& x^{1}&x^{49}&x^{61} & x^{65} &x^{99}& x^{ 117}& x^{153}& x^{186}\end{array}\right]\end{align*}
%We can use  a larger matrix $h_{GC}=\begin{bmatrix} M&I \end{bmatrix}$, for example, a submatrix of the  $7\times 127$ parity-check matrix for a Hamming $[127,120,3]$-code,   and obtain a code with an approximate rate of  $7/15$. 

We consider now the fully generalized case, where  $h_{GC}=\begin{bmatrix} M&I \end{bmatrix}$ for one constraint and $h'_{GC}=\begin{bmatrix} M_1&I&M_2 \end{bmatrix}$%\eqref{largech7}, 
obtained from $h_{GC}$ by rearranging the columns so that $I$ is in positions  8, 9, 10, and 11, to obtain  

\begin{align*}H_{GC}
&=\arraycolsep=2.54pt\left[\begin{array}{ccccccccccc|cccc} %0&1&2&3&4&5&6&7&&8 &9& 10&11&12&13&14\\\hline 
%1&x^{i_2}&x^{i_3}&x^{i_4}&x^{i_5}&x^{i_6}&x^{i_7}&x^{i_8}&x^{i_9}&x^{i_{10}}&x^{i_{11}}&x^{i_{12}}&x^{i_{13}}&x^{i_{14}}  &x^{i_{15}}\\\hline  
 1&0&x^{i_3}&0&x^{i_5}&0&x^{i_7}&x^{i_8}&0&0&0&x^{i_{12}}&0&x^{i_{14}}&x^{i_{15}}\\
 0&x^{i_2}&x^{i_3}&0&0& x^{i_6}&x^{i_7}& 0&x^{i_9}&0&0&0&x^{i_{13}}&x^{i_{14}}&x^{i_{15}}\\ 
  0&0&0&x^{i_4}& x^{i_5}&x^{i_6}&x^{i_7}& 0&0&x^{i_{10}}&0&x^{i_{12}}&x^{i_{13}}&x^{i_{14}}&0\\  
 1&x^{i_2}&x^{i_3}&x^{i_4}&x^{i_5}&x^{i_6}&x^{i_7}&0&0&0&x^{i_{11}}&0&0&0&0\\\hline
1&0&1&1& 1&0&1&0&1&0&1&1&0&0&0\\
0&1&1&1& 0&1&1&0&0& 1&1& 0&1&0&0\\ 
1&1&1&0&  0&0&0&1& 1&1&1& 0&0&1&0\\  
0&0&0&0& 1&1&1&1&1&1&1&0&0&0&1\end{array}\right]
\\&\sim \left[\begin{array}{c|c}  M_1^{\uparrow \{\begin{matrix}1& x^{i_2}&\cdots& x^{i_{11}}\end{matrix}\}}&M_2^{\uparrow \{\begin{matrix} x^{i_{12}}&\cdots&x^{i_{15}}\end{matrix}\}} \\\hline M&I\end{array}\right]  \sim \left[\begin{array}{c|c} H_{short}&0 \\\hline M&I\end{array}\right] ,\end{align*}
where \begin{align*}  H_{short}&=  M_1^{\uparrow \{\begin{matrix}1& x^{i_2}&\cdots& x^{i_m}\end{matrix}\}}  +M_2^{\uparrow \{\begin{matrix} x^{i_{m+1}}&\cdots&x^{i_{n_v}}\end{matrix}\}}M,%=\begin{bmatrix}\vect{f}_1 \\ \vect{f}_2\\ \vect{f}_3\\ \vect{f}_4 \end{bmatrix} , \text{ where } %\\&
%\left\{\begin{matrix}
%1+x^{i_{12}}+ x^{i_{14}}&x^{i_{13}}+x^{i_{14}}&x^{i_3}+x^{i_{12}}+x^{i_{13}}+x^{i_{14}}&x^{i_{12}}+x^{i_{13}}&x^{i_5}+ x^{i_{12}}+x^{i_{15}}&x^{i_7}++x^{i_{12}}+x^{i_{13}}+x^{i_{15}}&&&&&
%\vect{f}_1 &\defeq &\left[ \begin{array}{ccccccccccc} 1&0&x^{i_3}&0&x^{i_5}&0&x^{i_7}&x^{i_8}&0&0&0 \end{array}\right] -\begin{bmatrix} x^{i_{12}}&0&x^{i_{14}}&x^{i_{15}}\end{bmatrix}M \\\vect{f}_2&\defeq& \left[ \begin{array}{ccccccccccc} 0&x^{i_2}&x^{i_3}&0&0& x^{i_6}&x^{i_7}& 0&x^{i_9}&0&0  \end{array}\right]-\begin{bmatrix} 0&x^{i_{13}}&x^{i_{14}}&x^{i_{15}}\end{bmatrix} M \\\vect{f}_3&\defeq&\left[ \begin{array}{ccccccccccc} 0&0&0&x^{i_4}& x^{i_5}&x^{i_6}&x^{i_7}& 0&0&x^{i_{10}}&0 \end{array}\right] -\begin{bmatrix}x^{i_{12}}&x^{i_{13}}&x^{i_{14}}&0\end{bmatrix}M \\\vect{f}_4& \defeq &\left[ \begin{array}{ccccccccccc}
%1&x^{i_2}&x^{i_3}&x^{i_4}&x^{i_5}&x^{i_6}&x^{i_7}&0&0&0&x^{i_{11}}\end{array}\right] -\begin{bmatrix}0&0&0&0\end{bmatrix}M\end{matrix}\right.
\end{align*} and $G_{short}$ is a generator matrix for the code given by $ H_{short}$. Then 
Theorem~\ref{thm-directly} can be applied  to obtain  a generator matrix for the code given by $H_{GC}$.\end{example} 

\section{QC-GLDPC codes by observing Pre-lifting}\label{app:pre2}
\noindent{\bf Example \ref{example2-GLDPC} (cont.).} In this  example, we observe pre-lifting in the case  $n_v=7$, with $N=34$ and exponents $\begin{bmatrix}  1&x^{44}&x^{46}&x^{14}& x^{61}&x^{49}&x\end{bmatrix} $ that were used earlier. 
%We can improve the code parameters by applying the same idea as before,  rearranging  rows and columns in 
%
%\begin{align*}
%\begin{bmatrix} 1&1&1&1&1 & 1&1\\  1&x^{61}&x^{49}&x^{44}&x&x^{46}&x^{14}\end{bmatrix} 
%\end{align*}
%to display a sequence of two liftings (since $N$ is even), and obtain %~\eqref{largech2}. 
%\begin{align*} {\small\left[ \begin{array}{cc|cc|cc|cc|cc|cc|cc}%0&1&2&3&4&5&6&7&8 &9& 10&11\\\hline 
% 1 &0&1&0&1&0&1&0&1 &0& 1&0&1&0\\
%0&1&0&1&0&1&0&1 &0& 1&0 &1&0&1\\ \hline
%1&0& 0&x^{31}&0& x^{25}&x^{22}&0&0&x& x^{23} &0&x^{7}&0\\ 
%0&1& x^{30}&0& x^{24}&0&0&x^{22}&1&0& 0&x^{23} &0&x^{7} \end{array}\right]}.\end{align*}
%where the exponents are now in $\Z_{34}$. 
%\begin{figure*}
%\begin{align} \label{largech2}
%{\tiny\left[ \begin{array}{cc|cc|cc|cc|cc|cc|cc}%0&1&2&3&4&5&6&7&8 &9& 10&11\\\hline 
%1&0&1&0&1&0&1&0&1 &0& 1&0&1&0\\
%0&1&0&1&0&1&0&1 &0& 1&0 &1&0&1\\\hline 
%1&0& 0&x^{31}&0& x^{25}&x^{22}&0&0&x& x^{23} &0&x^{7}&0\\ 
%0&1& x^{30}&0& x^{24}&0&0&x^{22}&1&0& 0&x^{23} &0&x^{7}\end{array}\right]}\end{align}\hrulefill
%\end{figure*}
We first generalize  3 of the constraints with 
\begin{align*}
h_{GC, 0}\defeq \begin{bmatrix} M&I \end{bmatrix} \defeq \left[\begin{array}{cccc|ccc} 1&1&1&0&1&0&0\\1&1&0&1&0&1&0\\ 1&0&1&1&0&0&1\end{array}\right] \end{align*}
% If we rearrange the matrix so that all the even exponents come first, and then  lift to obtain 
%  \begin{align*} {\small\left[ \begin{array}{cc|cc|cc|cc|cc|cc|cc}
%% 0 &1   &2&    3&4    &5&6   &7&8    &9& 10 &11&  12 &13        \\\hline 
% 1 &0   &1&    0&1    &0&1   &0&1    &0& 1   &0&   1&0\\
%0&1&0&1&0&1&0&1 &0& 1&0 &1&0&1\\ \hline
%1&0&x^{22}&0& x^{23} &0&x^{7}&0 & 0&x^{31}&0& x^{25}&0&x\\ 
%0&1& 0&x^{22}& 0&x^{23} &0&x^{7}&x^{30}&0& x^{24}&0&1&0 \end{array}\right]}.\end{align*}
%and then generalize using only the matrix $h_{GC, 0}$
%%\begin{align*}h_{GC, 0}\defeq \begin{bmatrix} M&I \end{bmatrix} \defeq \left[\begin{array}{cccc|ccc} 1&1&1&0&1&0&0\\1&1&0&1&0&1&0\\ 1&0&1&1&0&0&1\end{array}\right], 
%%\end{align*}
% for the first three constrains, and leave the forth unchanged, we obtain: 
and obtain a rate $136/476=2/7$ code of minimum distance between 30 and 34. %, most likely 34.\footnote{We arranged these exponents to different positions within odds and evens, and it seems that 34 is the best among them. } 
%Because of this the matrix can be rearranged like in the example of $n=6$ as below, where we can group the last row in $\begin{bmatrix} \vect{0}& \begin{array}{|ccc|cccc|} x^{30}&x^{24}&1& 1& x^{22} & x^{23}&x^7\end{array} &\vect{0}\end{bmatrix} .$  
%
%\begin{align*}\left[\begin{array}{c||c}  
%\begin{matrix}M&I\\0&0\\M^{\uparrow\{0,22, 23, 7\}}&0 \\0& \begin{matrix}  x^{30}&x^{24}&1\end{matrix}\end{matrix} &~
% \begin{matrix}0&0\\M&I\\0&I^{\uparrow \{31, 25, 1\}}\\\begin{matrix} 1& x^{22} & x^{23}&x^7\end{matrix} &0\end{matrix}  \end{array}\right] 
%% \approx 
%%\left[\begin{array}{c|c} \begin{matrix}M&0 \end{matrix}&\begin{matrix} I&0\end{matrix} \\\begin{matrix}0&M\end{matrix}&\begin{matrix}0&I\end{matrix}\\\hline M^{\uparrow\{0,22, 23, 7\}}&  I^{\uparrow \{31, 25, 1\}}M\\\begin{bmatrix}   x^{30}&x^{24}&1\end{bmatrix}M & \begin{bmatrix} 1& x^{22} & x^{23}&x^7  \end{bmatrix} \end{array}\right]\\
% \end{align*}
% We can write the codewords in this equivalent code $\ker(H_{GC}')$ as vectors $\vect{v}\defeq (\vect{v}^\tr_1,\vect{v}^\tr_2,\vect{v}^\tr_3,\vect{v}^\tr_4)^\tr$ with $\vect{v}_i \defeq (v_{i,1},v_{i,2}, v_{i,3})^\tr\in \left(\Ftwo[x]/(x^{34}+1)\right)^3, i=2,4, \vect{v}_i \defeq (v_{i,1},v_{i,2}, v_{i,3}, v_{i,4})^\tr,  i=1,3,$ over $\Ftwo[x]/(x^{34}+1),$ such that 
%\begin{align*} &\vect{v}_2=M\vect{v}_1\\& \vect{v}_4=M\vect{v}_3\\
%& M^{\uparrow\{0,22, 23, 7\}}\vect{v}_1+ I^{\uparrow \{31, 25, 1\}}M\vect{v}_3=0\\
%&\begin{bmatrix}  x^{30}&x^{24}&1 \end{bmatrix} M\vect{v}_1+\begin{bmatrix} 1& x^{22} & x^{23}&x^7 \end{bmatrix}\vect{v}_3=0 \end{align*}
We compute $H_{1, short}$ as
%\begin{align*}
%h_{GC, 0}\defeq \begin{bmatrix} A&B \end{bmatrix} \defeq \left[\begin{array}{ccc|cccc} 1&1&1&0&1&0&0\\1&1&0&1&0&1&0\\ 1&0&1&1&0&0&1\end{array}\right], 
%\end{align*}
 \begin{align*} &H_{1, short} = 
\left[\begin{array}{c|c} M^{\uparrow\{\begin{matrix} 1&x^{22}&x^{23}&x^7\end{matrix}\}}&  I^{\uparrow \{\begin{matrix} x^{31}& x^{25} &x\end{matrix}\}}M\\\begin{bmatrix}   x^{30}&x^{24}&1\end{bmatrix}M & \begin{bmatrix} 1& x^{22} & x^{23}&x^7  \end{bmatrix} \end{array}\right]\\
=&\left[\begin{array}{ccc|ccccc} 
1& x^{22} &x^{23}&0&x^{31}&x^{31}& x^{31}& 0\\
1&x^{22}&0&x^7&x^{25}&x^{25}&0&x^{25}\\
1&0&x^{23}&x^7& x& 0&x&x \\\hline
x^{30}+x^{24}+1& x^{30}+x^{24}&1+x^{30}&1+x^{24} &1& x^{22} & x^{23}&x^7 \end{array} \right] \\&
=\left[\begin{array}{c|c} A&B\\\hline \begin{bmatrix} x^{30}+x^{24}+1& x^{30}+x^{24}&1+x^{30}\end{bmatrix}& \begin{bmatrix}1+x^{24}& 1& x^{22} & x^{23}&x^7\end{bmatrix} \end{array} \right] \\&\sim \left[\begin{array}{c|c}I & A^{-1} B\\\hline\begin{bmatrix}x^{30}+x^{24}+1& x^{30}+x^{24}&1+x^{30}\end{bmatrix} & \begin{bmatrix}1+x^{24} & 1& x^{22} & x^{23}&x^7  \end{bmatrix} \end{array}\right]\\&
 \sim \left[\begin{array}{c|c} I & A^{-1} B\\\hline\vect{0}& \begin{bmatrix}1+x^{24} & 1& x^{22} & x^{23}&x^7  \end{bmatrix} -\begin{bmatrix} x^{30}+x^{24}+1& x^{30}+x^{24}&1+x^{30}\end{bmatrix}A^{-1} B\end{array}\right],\end{align*} where \begin{align*}
 & A^{-1} =\begin{bmatrix} 1&1&1\\x^{-22}&0&x^{-22}\\
 x^{-23}&x^{-23}&0\end{bmatrix},\end{align*}\begin{align*} H_{2, short}\defeq &\begin{bmatrix}1+x^{24}&1& x^{22} & x^{23}&x^7  \end{bmatrix} -\begin{bmatrix} x^{30}+x^{24}+1& x^{30}+x^{24}&1+x^{30}\end{bmatrix}A^{-1} B\\=&
 \begin{bmatrix} f_1&f_2&f_3&f_4&f_5\end{bmatrix},\end{align*}
 
 \pagebreak
 \begin{align*}
& f_1=x^{24 } + x^{18 } + x^{15 } + x^{14 } + x^9 + 1, \\& f_2=x^{33 } + x^{32 } + x^{27 } + x^{15 } + x^{9 } + x^8 + x^5 + x^4 + x^3 + x^2 + x + 1, \\&f_3=x^{33 } + x^{32 } + x^{31 } + x^{27 } + x^{25 } + x^{22 } + x^{15 } + 
    x^{8 } + x^{5 } + x^{4 } + x^{2},\\&f_4=  x^{33 } + x^{27 } + x^{25 } + x^{23 } + x^{21 } + x^{9 } + x^{8 } + x^{5 } + 
    x^{4 } + x^{3 } + x,\text{ and}\\& f_5= x^{32 } + x^{31 } + x^{21 } + x^{15 } + x^{9 } + x^{7 } + x^{3 } + x^{2 } + x.\end{align*}
  The code $\ker(H_{1, short})$  is a $[272, 136, 18]$  code, while $\ker(H_{2, short})$ is a $[170, 136, 8]$ code 
  with $ (0, x^{12}+ x^{31}, x, x^3+x^{17}+x^{18},x^{14}+x^{16})$ being a codeword of weight 8. 
%  (0 0 0 0 0 0 0 0 0 0   
%   0 0 0 0 0 0 0 0 0 0 
%   0 0 0 0 0 0 0 0 0 0 
%   0 0 0 0 
%   0 0 0 0 0 0 0 0 0 0 
%   0 0 1 0 0 0 0 0 0 0 
%   0 0 0 0 0 0 0 0 0 0 
%   0 1 0 0 
%   0 1 0 0 0 0 0 0 0 0 
%   0 0 0 0 0 0 0 0 0 0 
%   0 0 0 0 0 0 0 0 0 0 
%   0 0 0 0 
%   0 0 0 1 0 0 0 0 0 0 
%   0 0 0 0 0 0 0 1 1 0 
%   0 0 0 0 0 0 0 0 0 0 
%   0 0 0 0 
%   0 0 0 0 0 0 0 0 0 0 
%   0 0 0 0 1 0 1 0 0 0 
%   0 0 0 0 0 0 0 0 0 0 
%   0 0 0 0)   $
We can use these small weight codewords to create codewords in $\ker(H_{1, short})$, and then in $\ker(H_{GC})$, thus obtaining upper bounds and lower bounds for the original code. 
   
Using Theorem~\ref{thm-directly},  a generator matrix $G_{2,short}$ of the form \eqref{Gshort}, for   $\ker(H_{2, short})$ leads to a generator matrix  $\begin{bmatrix} G_{2,short}(A^{-1}B)^\tr&G_{2,short}\end{bmatrix}$ for $\ker(H_{1, short})$ which, in turn, leads to a generator matrix for the original code,\footnote{We observe that the $ k\times  5$ matrix $G_{2,short}$,  with $\rank (G_{2,short} ) =k$,    and the $3\times 4$ matrix $M^\tr$ can not be multiplied, but we can multiply  $\begin{bmatrix} G_{2,short}(A^{-1}B)&G_{2,short}\end{bmatrix} \diag (M^\tr, M^\tr)$. Indeed, 
 $G_{2,short}(A^{-1}B)^\tr$ is a  $k \times  3$  matrix, which gives the matrix $\begin{bmatrix} G_{2,short}(A^{-1}B)^\tr&G_{2,short}\end{bmatrix} $, a $k \times 8$ matrix.  The matrix $\diag (M^\tr, M^\tr)$ is an $8\times 6$ matrix, giving that the product \\$\begin{bmatrix} G_{2,short}(A^{-1}B)&G_{2,short}\end{bmatrix} \diag (M^\tr, M^\tr)$ is a $k\times 6$ matrix. Finally, $G$ is a $k\times 14$ matrix, as it should be. }
 $$G=\begin{bmatrix} G_{2,short}(A^{-1}B)^\tr&G_{2,short}& \begin{bmatrix} G_{2,short}(A^{-1}B)&G_{2,short}\end{bmatrix} \diag (M^\tr, M^\tr) \end{bmatrix}.
 $$We can also  generalize the same matrix  using  $h_{GC, 0}\defeq \begin{bmatrix} M&I \end{bmatrix}  $ for the first two constraints, $h'_{GC, 0}\defeq \begin{bmatrix} I&M \end{bmatrix}  $ for the third block of constraints, and leave the last unchanged, to obtain
%
%\begin{align*}{\scriptsize \left[ \begin{array}{cc|cc|cc|cc|cc|cc|cc}
%0&1&2&3&4&5&6&7&8 &9& 10&11&12&13\\\hline 
%1&0&1& 0&1&0&0&0&1 &0& 0&0&0&0\\
%1&0&1&0&0&0&1&0&0 &0& 1&0&0&0\\
%1&0&0&0&1&0&1&0&0&0&0&0& 1&0\\\hline 
%0&1&0&1&0&1&0&0&0&1 &0& 0&0&0\\
%0&1&0&1&0&0&0&1&0&0 &0& 1&0&0\\
%0&1&0&0&0&1&0&1&0&0&0&0&0& 1\\\hline 
%1&0&  0&  0&0& 0& x^{7}&   0&  0&          x^{31}&0&   x^{25}&  0&0 \\
%0&0&  x^{22}&0&  0&0&    x^{7}& 0  &0&   x^{31}& 0&0&  0 &x\\ 
%0&0&  0&  0&x^{23}&   0&   x^{7}&0&0&   0&          0&    x^{25}   &   0   &x\\\hline 
%0&1& 0&x^{22}& 0&x^{23} &0&x^{7}&x^{30}&0& x^{24}&0&1&0\end{array}\right]}
%\end{align*} %~\eqref{largech3} 
 a rate $136/476=2/7$ 
code of the  minimum distance between 24 and 27. 
%It is larger than the one above, when we considered two matrices. 

%I arranged the exponents to first be even and correspond to the generalizing  by $M$ 
%and the last three odd, corresponding to generalizing by $I$. Because of this the matrix can be rearranged like in the example of $n=6$ as below, where we can group the last row in $\begin{bmatrix} \vect{0}& \begin{array}{|ccc|cccc|} x^{30}&x^{24}&1& 1& x^{22} & x^{23}&x^7\end{array} &\vect{0}\end{bmatrix} .$  
We set $$J\defeq \begin{bmatrix} 1&0&0&1 \\0&1&0&1\\0&0&1&1 \end{bmatrix}, P\defeq \begin{bmatrix} 1&1&0 \\1&0&1\\0&1&1 \end{bmatrix}, \text{ with }h_{GC, 1} =\begin{bmatrix} J&P\end{bmatrix},$$
%\begin{align*} 
%\left[\begin{array}{c|c|c|c}
%M&I&0&0\\
%0&0&M&I\\
% J^{\uparrow\{0,22, 23, 7\}}  &0 &0&P^{\uparrow \{31, 25, 1\}}\\\hline0& \begin{matrix}  x^{30}&x^{24}&1\end{matrix}&\begin{matrix} 1& x^{22} & x^{23}&x^7\end{matrix} &0\end{array}\right] \end{align*}
% We can write the codewords in this equivalent code $\ker(H_{GC}')$ as vectors $\vect{v}\defeq (\vect{v}^\tr_1,\vect{v}^\tr_2,\vect{v}^\tr_3,\vect{v}^\tr_4)^\tr$ with $\vect{v}_i \defeq (v_{i,1},v_{i,2}, v_{i,3})^\tr\in \left(\Ftwo[x]/(x^{34}+1)\right)^3, i=2,4, \vect{v}_i \defeq (v_{i,1},v_{i,2}, v_{i,3}, v_{i,4})^\tr,  i=1,3,$ over $\Ftwo[x]/(x^{34}+1),$ such that 
%\begin{align*} &\vect{v}_2=M\vect{v}_1\\& \vect{v}_4=M\vect{v}_3\\
%& J^{\uparrow\{0,22, 23, 7\}}\vect{v}_1+ P^{\uparrow \{31, 25, 1\}}M\vect{v}_3=0\\
%&\begin{bmatrix}  x^{30}&x^{24}&1 \end{bmatrix} M\vect{v}_1+\begin{bmatrix} 1& x^{22} & x^{23}&x^7 \end{bmatrix}\vect{v}_3=0 \end{align*}
then, 
 \begin{align*} &H_{1, short} \defeq 
\left[\begin{array}{c|c} J^{\uparrow\{\begin{matrix}1& x^{22}& x^{23}& x^7\end{matrix} \}}&  P^{\uparrow \{\begin{matrix}x^{31}& x^{25} &x\end{matrix} \}}M\\\begin{bmatrix}   x^{30}&x^{24}&1\end{bmatrix}M & \begin{bmatrix} 1& x^{22} & x^{23}&x^7  \end{bmatrix} \end{array}\right]=\\
&\left[\begin{array}{ccc|ccccc} 
1& 0 &0&x^7&x^{31}&x^{31}& x^{31}& 0\\
0&x^{22}&0&x^7&x^{25}&x^{25}&0&x^{25}\\
0&0&x^{23}&x^7& x& 0&x&x \\\hline
x^{30}+x^{24}+1& x^{30}+x^{24}&1+x^{30}&1+x^{24} &1& x^{22} & x^{23}&x^7 \end{array}\right]\sim \\&
\left[\begin{array}{ccc|ccccc} 
1& 0 &0&x^7&x^{31}&x^{31}& x^{31}& 0\\
0&1&0&x^{19}&x^{3}&x^{3}&0&x^{3}\\
0&0&1&x^{18}& x^{12}& 0&x^{12}&x^{12} \\\hline
x^{30}+x^{24}+1& x^{30}+x^{24}&1+x^{30}&1+x^{24} &1& x^{22} & x^{23}&x^7 \end{array}\right]\sim\\&\left[\begin{array}{c|c} I&A\\\hline0&\begin{matrix} f_1& f_2&f_3&f_4&f_5\end{matrix}
 \end{array}\right], \text{ where } A\defeq \begin{bmatrix}x^7&x^{31}&x^{31}& x^{31}& 0\\x^{19}&x^{3}&x^{3}&0&x^{3}\\ x^{18}& x^{12}& 0&x^{12}&x^{12} \end{bmatrix}\text{ and} \\&\begin{bmatrix} f_1& f_2&f_3&f_4&f_5\end{bmatrix} =\begin{bmatrix}1+x^{24} &1& x^{22} & x^{23}&x^7\end{bmatrix}-\begin{bmatrix}x^{30}+x^{24}+1& x^{30}+x^{24}&1+x^{30} \end{bmatrix}A.\end{align*}
 From this form, we can proceed using \eqref{Gshort} as before to find a generator matrix. 
 The minimum distance  of $\ker(H_{1, short})$ is a lower bound for the minimum distance of the code, and a generator matrix of this code will lead to a generator matrix for the larger code. \hfill $\Box$ 

\end{document}